\documentclass[prb,twocolumn,superscriptaddress,showpacs,amsmath,amssymb]{revtex4}

\usepackage{graphicx}
\usepackage{latexsym}
\usepackage{amsmath}
\usepackage{amssymb}
\usepackage{amsfonts}
\usepackage{color}

\usepackage{amssymb}
\usepackage{bm}
\usepackage{verbatim}

\usepackage[colorlinks,bookmarks=false,citecolor=blue,linkcolor=red,urlcolor=blue]{hyperref}
\usepackage{color}
\usepackage[all]{hypcap} 

\begin{document}


\title{Topological transitions in electronic spectra:\\
Crossover between Abrikosov and Josephson vortices}

\author{A.~V.~Samokhvalov}
\affiliation{Institute for Physics of Microstructures, Russian
Academy of Sciences, 603950 Nizhny Novgorod, GSP-105, Russia}
\affiliation{Lobachevsky State University of Nizhni Novgorod,
603950 Nizhni Novgorod, Russia}
\author{V.~D.~Plastovets}
\affiliation{Institute for Physics of Microstructures, Russian
Academy of Sciences, 603950 Nizhny Novgorod, GSP-105, Russia}
\affiliation{Lobachevsky State University of Nizhni Novgorod,
603950 Nizhni Novgorod, Russia}
\affiliation{ Sirius University of
Science and Technology, 1 Olympic Ave, 354340 Sochi, Russia}
\author{A.~S.~Mel'nikov}
\affiliation{Institute for Physics of Microstructures, Russian
Academy of Sciences, 603950 Nizhny Novgorod, GSP-105, Russia}
\affiliation{Lobachevsky State University of Nizhni Novgorod,
603950 Nizhni Novgorod, Russia}
\affiliation{
Sirius University of Science and Technology, 1 Olympic Ave, 354340 Sochi, Russia}

\date{\today}
\begin{abstract}
The electronic structure of a vortex line trapped by a planar
defect in a type-II superconductor is analyzed within the
Bogoliubov-de Gennes theory. The normal reflection of electrons
and holes at the defect plane results in the topological
transition in the spectrum and formation of a new type of
quasiparticle states skipping or gliding along the defect. This
topological transition appears to be a hallmark of the initial
stage of the crossover from the Abrikosov to the Josephson vortex
type revealing in the specific behavior of the quantized
quasiparticle levels and density of states. The increase in the
resulting hard and soft gaps affects the vortex mobility along the
defect plane and splitting of the zero bias anomaly in the
tunneling spectral characteristics.
\end{abstract}


\maketitle

\section{Introduction}\label{Intro}

\begin{figure}[t!]
\includegraphics[width=0.25\textwidth]{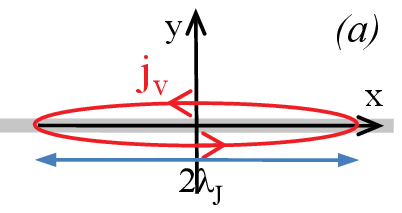}
\includegraphics[width=0.25\textwidth]{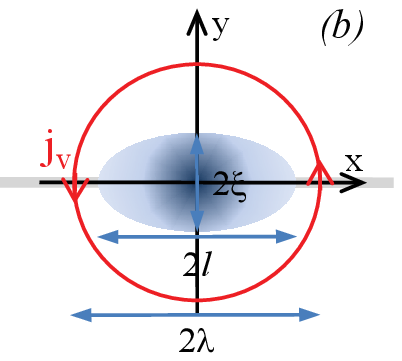}
\includegraphics[width=0.25\textwidth]{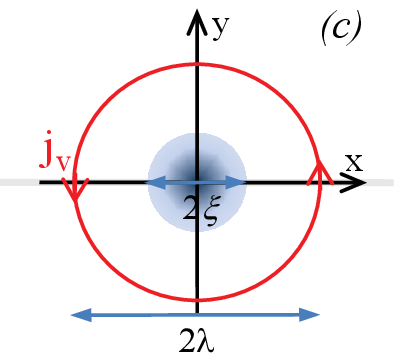}
\caption{(Color online) Vortex pinned by a planar defect
positioned in the $y = 0$ plane for several values of the barrier
transparency $D$: ($a$) $D \ll D_{\lambda}$ -- Josephson vortex
for weak coupling ($l \gg \lambda_J \gg \lambda$); ($b$) $D \sim
D_{\lambda}$ -- Abrikosov-like vortex for strong coupling ($l \sim
\lambda_J \sim \lambda$); ($c$) $D \sim D_{\xi}$ -- Abrikosov
vortex ($l \sim \xi$). The region of the vortex core is shown by
grey color. Current streamline around the vortex is shown by a
solid red line.} \label{Fig:1}
\end{figure}

The most general definition of different vortex type solutions for
the order parameter in superconducting and superfluid systems is
based on the calculation of the so-called circulation of the
gradient of the  order parameter phase around the line of
singularity. Provided this circulation equals to $2\pi$ we get a
singly quantized vortex. The particular structure of the order
parameter and magnetic field  distributions strongly depends then
on the specific system. In a homogeneous isotropic superconductor
the vortex solution  possessing a cylindrical symmetry  is well
known as an Abrikosov vortex \cite{Abrikosov57} while the presence
of any anisotropy or inhomogeneity can strongly deform this vortex
line in the plane perpendicular to its axis (see Fig.1). An
extreme example of such anisotropic solution which does not even
possess the normal core can be realized for a vortex pinned at the
Josephson junction \cite{Barone-Josephson}. Such quasi-one
dimensional vortices are also called Josephson vortices (see
Fig.~\ref{Fig:1}$a$) and are known to play an important role in
magnetic and transport properties of layered and nano-structured
systems. Provided the junction critical current density $j_c$ is
much smaller than depairing current density
\begin{equation}\label{eq:lambda_d}
    j_d = c \Phi_0 / 12 \sqrt{3} \pi^2 \lambda^2 \xi\,,
\end{equation}
the Josephson penetration depth
\begin{equation}\label{eq:lambda_J}
    \lambda_J = \sqrt{c \Phi_0 / 16 \pi^2 j_c \lambda}\,,
\end{equation}
appears to be much larger than the London penetration depth
$\lambda$.
Here $\Phi_0=\pi \hbar c / e$ is the magnetic flux quantum, and
$\xi$ is the superconducting coherence length.
Clearly, changing the electron transparency of the junction one
can get a variety of intermediate vortex states corresponding to a
crossover from the Josephson to the Abrikosov vortex
\cite{Gurevich-PRB92,Horide-PRB07,Horide-PRB08}. This situation
with the intermediate transparencies naturally appears in many
superconducting systems studied in experiments, e.g. in
superconductors with twinning planes
\cite{Khlyustikov-Buzdin-AdvPh87}, low-angle grain boundaries
\cite{Gurevich-PRL02,Hilgenkamp-RMP02} or other types of defects
\cite{Jooss-PRB00, Djupmyr-PRB05,Tafuri1-RPP05}. An appropriate
theoretical treatment needed, for instance, for the interpretation
of the experimental data on the magnetic field distribution can be
well developed on the basis of the Ginzburg--Landau theory.
Indeed, using a general expression \cite{Golubov-RMP04} for the
critical current $I_c$ across the junction with a cross-section
area $S$
\begin{equation}\label{eq:I_c}
    I_c = j_c S= \pi \Delta_0 / 2 e R_N \,,
\end{equation}
and relation between the contact resistance and the angle-averaged transmission probability of the
barrier $D$
\begin{equation}\label{eq:R_N}
    R_N^{-1} = k_F^2 S\, ( 2 e^2 / \hbar )\, D  \,,
\end{equation}
we derive the following
simple relation
\begin{equation}\label{eq:lambda_J2}
    \lambda_J^2 = \lambda\, \xi / 12 \pi^2 D \,.
\end{equation}
It is natural that the Josephson length $\lambda_J$ grows if the
transmission probability of the barrier $D$ decreases. To satisfy
the relation $\lambda_J \gg \lambda$, the barrier transparency
should be small enough: $D \ll D_{\lambda} = 1 / 12 \pi^2 \kappa
\ll 1$, where $\kappa = \lambda / \xi$ is the Ginzburg--Landau
parameter. As the probability of electron transmission through the
barrier grows above $D_{\lambda}$ the changes in the structure of
the order parameter are controlled by the relation between the
Josephson length $\lambda_J$, the London penetration depth
$\lambda$ and the coherence length $\xi$. Keeping in mind type-II
superconductors we should take
 $\xi \lesssim \lambda$.
When the current density $j(r)$ in the vortex core ($r \lesssim \xi$)
becomes of order of the depairing one $j_d$, the length $l$ of the core
along the defect can be estimated from the continuity of currents
flowing parallel and perpendicular to the defect within the core \cite{Gurevich-PRB92}:
$l\,j_c \sim j_d \xi$, whence
$
    l \sim j_d \xi / j_c \sim \lambda_J^2 / \lambda \,.
$
The case $D \gtrsim D_{\lambda}$ ($\xi <  l \lesssim \lambda
\sim \lambda_J$) corresponds to the limit of strong Josephson coupling with $j_c
\gtrsim j_d / \kappa$ , and we can no more consider the solution in the form of
 a core free Josephson vortex having the size of the order $\lambda_J$.
Instead, we get the crossover to the
Abrikosov-like vortex having strongly deformed anisotropic core
($l \times \xi$), where the superconducting order parameter is
suppressed (see Fig.~\ref{Fig:1}$b$). The distributions of
the magnetic field and circular screening currents outside the
core ($r \gg l,\, \xi$) approach now with the ones for the Abrikosov vortex in a
uniform superconductor. In the case of the extremely strong Josephson
coupling $D \gtrsim D_{\xi} = 1 / 12 \pi^2$ ($l \lesssim \xi$) the
anisotropy of the vortex core becomes negligible, and
at this initial stage of the crossover (see Fig.~\ref{Fig:1}$c$)
the order parameter profile in the Abrikosov vortex core is almost insensitive to the defect.
%
%

Despite general correctness of the above qualitative picture
 there exist several important physical issues which definitely can not be
described within the phenomenological model and demand a more
careful microscopic consideration. This statement surely relates
to the scanning tunneling microscopy (STM) and spectroscopy (STS)
data which provide detailed spatially resolved excitation spectra
\cite{Hess-PRL89,Hoogenboom-PRB00,Guillamon-PRL08,Karapetrov-PRL05,Roditchev-NatPh15}
and also to the problem of the vortex dynamics and dissipation
\cite{Kopnin-TNSC,Guinea-Pogorelov-PRL95,Feigelman-Skvortsov-PRL97,%
Larkin-Ovchinnikov-PRB98,Skvortsov-PRB03, Melnikov-AVS-JETPL11}.
In the latter case the crossover from the Abrikosov to the
Josephson vortex is particularly important since it is accompanied by
the disappearance of the normal vortex core which provides the
dominating contribution to the dissipation and resulting vortex
viscosity \cite{Gurevich-PRL02}.

\begin{figure}[t!]
\includegraphics[width=0.4\textwidth]{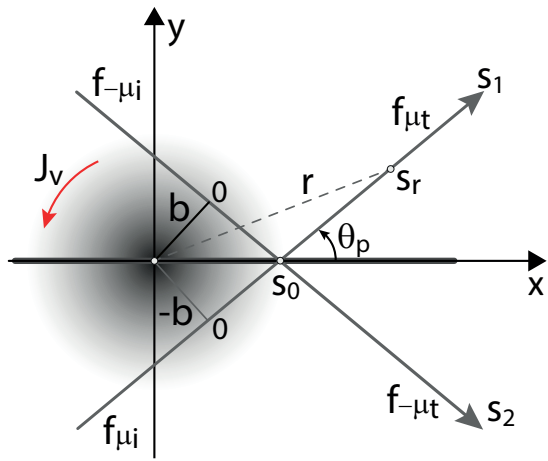}
\caption{(Color online) Specular reflection of quasiclassical
trajectories $s_1$ and $s_2$ with opposite values of the angular
momentum $\mu = \pm k_{\bot}\, | b |$ at the defect in the plane
$y=0$. The region of the vortex core is marked by the gray color.
The red arrow shows the direction of the supercurrent in the
vortex.} \label{Fig:2}
\end{figure}

Considering the microscopic theory one should take into account
the behavior of the subgap fermionic states bound to the Abrikosov
vortex core which are known to determine both the structure and
dynamics of vortex lines in the low temperature limit
\cite{Kopnin-TNSC}. These subgap states are known to form the
so--called anomalous spectral branch crossing the Fermi level. For
well separated vortices the behavior of the anomalous branches can
be described by the Caroli--de Gennes--Matricon (CdGM) theory
\cite{Caroli-PL64}: for each individual vortex the energy
$\varepsilon_{CdGM}(\mu)$ of subgap states varies from $-\Delta_0$
to $+\Delta_0$ as one changes the angular momentum $\mu$ defined
with respect to the vortex axis. Here $\Delta_0$ is the
superconducting gap value far from the vortex axis. At small
energies $|\varepsilon|\ll\Delta_0$ the spectrum is a linear
function of $\mu$: $\varepsilon_{CdGM}(\mu) \simeq -\mu \hbar
\omega_0$, where $\hbar \omega_0\approx \Delta_0 /(k_F \xi) =
\Delta_0^2 / 2 E_F \ll \Delta_0$ is the interlevel spacing, $\xi =
\hbar V_F/\Delta_0$, $k_F$, $V_F$ and $E_F$ are the Fermi
momentum, velocity and energy, respectively. Neglecting the
quantization of the angular momentum $\mu$ one can get the
anomalous spectral branch crossing the Fermi level at $\mu=0$ for
all orientations of the momentum $\mathbf{k}_F = k_F\left(
\cos\theta_p,\, \sin\theta_p \right)$. Thus, in the space
($\mu-\mathbf{k}_F$) we obtain a Fermi surface (FS) for
excitations localized within the vortex core (see
Ref.~\onlinecite{Volovik-Universe} for review). For a fixed energy
$\varepsilon$ we can define a quasiclassical orbit in the plane
($\mu-\theta_p$): $\mu(\theta_p) = - \varepsilon / \hbar
\omega_0$. Each point at this orbit corresponds to a straight
trajectory passing through the vortex core (Fig.~\ref{Fig:2}). The
precession of quasiparticle trajectory along the orbit is
described by the Hamilton equation: $\hbar
\partial\theta_p / \partial t =
\partial \varepsilon / \partial\mu$.

The wave functions of the subgap states are localized inside the
vortex core because of the Andreev reflection of quasiparticles at
the core boundary. Any additional normal scattering process should
modify the behavior of the anomalous spectral branch. Such
modification can be noticeable even for impurity atoms introduced
in a vortex core \cite{Larkin-PRB98} and becomes much more
pronounced provided we consider a vortex pinned by a normal-metal
\cite{Tanaka-JJAP95,Eschrig} or an insulating
\cite{Melnikov-Samokhvalov-PRB09,Rosenstein-PRB11_microwave-absorp,%
Melnikov-Samokhvalov-JETPL15_review,Vadimov-Melnikov-JLTP16_chiral}
columnar defect of the size $R \ll \xi$ well exceeding the Fermi
wavelength. In the last case the scattering at the defect is
responsible for the opening of the minigap $\varepsilon_0 \sim
\Delta_0 R / \xi$ in the spectrum of localized states and
resulting suppression of the dissipation at low temperatures $T
\ll \varepsilon_0$ \cite{Kopnin-TNSC,LL-IX-2}.  For a vortex
approaching a flat or curved sample boundary an appropriate
spectrum transformation was studied in
Refs.~\cite{Kopnin-PRL05,Kopnin-PRB07,Melnikov-PRB08,Melnikov-PRB09-LDOS-mesa}.
Change in the anomalous spectral branch is accompanied by the
changes in  the topology  of quasiclassical orbits in the
($\mu-\theta_p$) plane. Such topological transitions in
quasiparticle spectra of vortex systems are similar to the
well--known Lifshits transitions which occur in the band spectra
of metals \cite{lifshits,blanter}. The generic examples of such
transitions in vortex matter including the opening of the closed
segments of the orbits in the ($\mu-\theta_p$) plane or merging
and reconnection of the different segments via the Landau-Zener
tunneling have been previously studied in Refs.
\cite{Melnikov-PRB08,Volovik-gapless,mel-silaev}. The basic
properties of vortex matter such as pinning and transport
characteristics, heat transport in the vortex state and
peculiarities of the local density of states should be strongly
affected by these changes in the topology of the subgap spectral
branches.

%
%
It is the goal of the present work to develop a theoretical
description of the changes in the electronic structure of the pinned
vortex core which occur during the crossover between the Abrikosov and Josephson
vortices and unveil a
nontrivial topological nature of this vortex core transformation.
We restrict ourselves to situations when the barrier is rather
weak assuming $D \gtrsim D_{\xi}$, and focus on the modification
of the anomalous energy branches which occurs in a vortex pinned
by a planar defect  due to the quasiparticle normal reflection
at the defect boundary.

To elucidate our main findings we start from the simplified qualitative picture
illustrating the effect of the barrier on the quasiparticle subgap states.
First, considering the specular reflection of the quasiclassical trajectories
at the plane defect in Fig.~(\ref{Fig:2}) one can clearly see that the scattering couples the
wavefunctions with the opposite angular momenta $\pm \mu$.
Phenomenologically one can describe this coupling by a standard two-level problem:
$$
    \left( \varepsilon - \varepsilon_{\mu} \right)
    \left( \varepsilon - \varepsilon_{-\mu} \right) \approx
    \left( V_{gap} (\theta_p) \right)^2\,,
$$
where $\varepsilon_{\mu}$ denotes the anomalous spectral branch
for a linear trajectory passing through the core of a free vortex.
The scattering obviously can not couple the trajectories with
$\theta_p = 0\,, \pm \pi$, which are parallel to the defect plane.
Considering now the limit of small angles $\theta_p$ one can
expect that even for the barriers with rather good transparency
the tunneling probability should vanish in this angular interval.
The splitting of the energy levels around $\varepsilon =0$ should
originate from the superconducting phase difference at the ends of
the incident and reflected trajectories. This phase difference
equals to $\pi-2\theta_p$. Using now a standard expression for the
subgap Andreev state energy in a one-dimensional Josephson
junction \cite{Beenakker} we find: $\varepsilon=\pm
\Delta_0\cos(\pi/2-\theta_p)\simeq \pm \Delta_0\theta_p$. This
energy splitting gives us the estimate for the coupling
coefficient in the above two-level problem: $V_{gap} \sim
\Delta_0\theta_p$.

\begin{figure}[t!]
\includegraphics[width=0.4\textwidth]{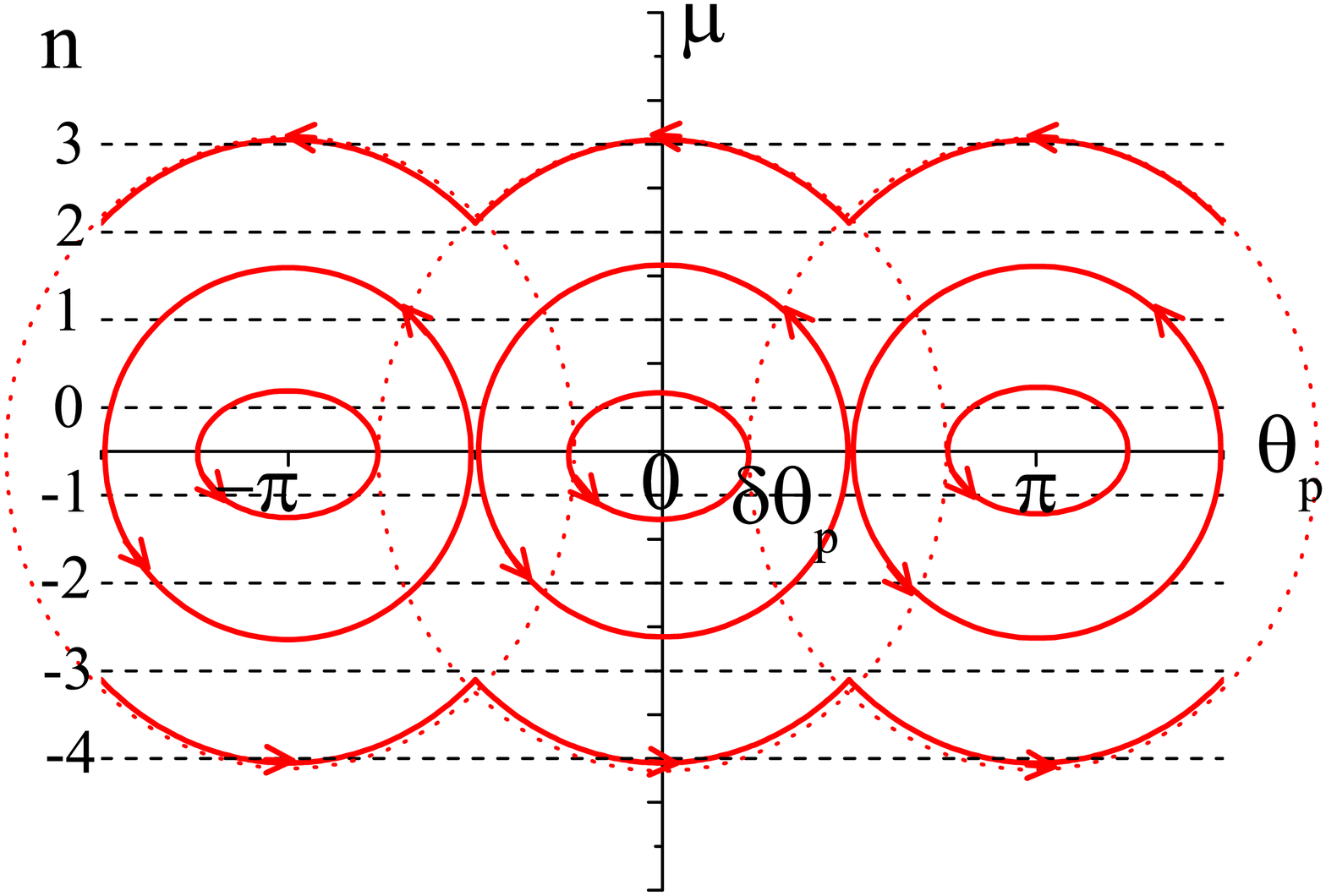}
\caption{(Color online) Quasiparticle orbits (\ref{eq:Orbits}) in
the ($\mu-\theta_p$) plane corresponding to different energy
levels $n$ are shown schematically by red solid lines. For
reference dashed lines show the orbits for a single Abrikosov
vortex in absence of a barrier. Arrows show the direction of the
quasiparticle trajectory precession along the orbit.}
\label{Fig:3}
\end{figure}
%
As a result, one obtains a set of quasiclassical orbits in
($\mu-\theta_p$) space
\begin{equation}\label{eq:Orbits}
    \mu(\theta_p) = \pm \frac{1}{\hbar \omega_0} \sqrt{\varepsilon^2 - \Delta_0^2
    \theta_p^2}\,.
\end{equation}
These orbits (\ref{eq:Orbits}) corresponding to the precession of
the quasiparticle trajectory are schematically shown in
Fig.~\ref{Fig:3}. For low energy levels one can clearly observe
the formation of closed orbits near the points $\theta_p = 0,\,
\pm \pi$, which are  separated by the prohibited angular domains
centered at $\theta_p = \pm \pi /2$. The closed orbits are nothing
more but skipping (or gliding) quasiparticle states formed due to
the scattering at the defect plane. The discrete subgap energy
levels of quasiparticles can be obtained from the semiclassical
Bohr-Sommerfeld quantization rule for canonically conjugate
variables $\mu$ and $\theta_p$
\cite{Kopnin-JETPL96-BohrSomm,Kopnin-PRL97-BohrSomm}:
\begin{equation}\label{eq:Bohr-Somm}
    \Sigma(\varepsilon) = \int\limits_0^{2\pi} \mu(\varepsilon,\theta_p) d \theta_p = 2\pi (n + \beta)\,,
\end{equation}
where $n$ is integer, $2 \pi$ is the period of $\mu(\theta_p)$,
and $\beta$ is of the order unity. Applying the Bohr-Sommerfeld
rule (\ref{eq:Bohr-Somm}) to the closed paths in ($\mu-\theta_p$)
space, we obtain the spectrum in the form
\begin{equation}\label{eq:Spacing0}
    \varepsilon_n^2  = \frac{\Delta_0^3}{E_F} (n + \beta) \,,
\end{equation}
which is dramatically different from the CdGM spectrum
$\varepsilon_n = \hbar \omega_0 (n + 1/2)$ and reminds the
square-root quantization of the quasiparticle spectra in different
types of nodal problems (like graphene \cite{graphene1,graphene2}
or d-wave superconductors in magnetic fields \cite{Janko-prl99}).
The novel minigap $\varepsilon_0 \simeq \Delta_0 \sqrt{\Delta_0 /
E_F}$ determined by Eq.~\ref{eq:Spacing0} well exceeds the CdGM
interlevel spacing $\hbar \omega_0$. This minigap increase
obviously manifests the partial suppression of the spectral flow
which should give the origin to all the dissipation phenomena
inside the vortex core during its motion. In this sense this
spectrum change can be viewed as a precursor to the crossover to
the Josephson vortex where all the subgap quasiparticle levels are
repelled from the Fermi energy to the gap value $\Delta_0$. On the
other hand, the limit of the moderate barrier strength studied
here provides a possibility to observe a novel type of the vortex
core with the peculiar quantization rule arising from the
splitting of the orbit segments in the $\mu-\theta_p$ plane. This
splitting destroys the trajectory precession in the whole angular
interval $0<\theta_p<2\pi$ changing, thus, the topology of the
quasiclassical orbits. The precession region $|\theta_p| \le
\delta\theta_p$ expands with an increase of the energy level $n$.
As a result, for rather high levels the prohibited angular domains
shrink, the precession over the full region $0 \le \theta_p \le
2\pi$ restores, and we get the crossover to a CdGM type of
spectrum $\varepsilon_n \sim n$.

The paper is organized as follows. In
Sec.~\ref{EqnSection} we introduce the basic equations used
for the spectrum calculation. In Sec.~\ref{SingleQuant} we study
the quasiparticle spectrum transformation for a singly quantized
vortex pinned at the planar defect and discuss the consequencies for the vortex dynamics.
 The Sec.~\ref{LocCond} is
devoted to the analysis of the peculiarities of the local density
of states for a vortex pinned at the defect. We summarize our
results in Sec.~\ref{sum}.

\section{Basic equations}\label{EqnSection}

Hereafter we consider a planar defect in the plane $y=0$ as a
$\delta-$function repulsive potential for quasiparticles, i.e.
$V(y) = H \delta(y)$. The magnetic field $\mathbf{B} = B
\mathbf{z}_0$ is assumed to create a single quantum vortex line
parallel to the $z-$axis trapped inside the attractive potential
well within the defect \cite{Blatter-RMP94}. The vortex center
defined as a point of the order parameter phase singularity is
positioned at the point $x = y = 0$.

We assume the system to be homogeneous along the $z-$axis, thus,
the $k_z-$projection of the momentum is conserved. The quantum
mechanics of quasiparticle excitations in a superconductor is
governed by the two dimensional BdG equations for particlelike
($u$) and holelike ($v$) parts of the two-component quasiparticle
wave functions $\hat{\Psi}(\mathbf{r},z)=(u,\,v)\,\exp(i k_z z)$:
\begin{subequations}\label{eq:BdG}
\begin{eqnarray}
 -\frac{\hbar^2}{2m} \left( \nabla^2 + k_{\bot}^2 \right)\,u
      + \Delta(\mathbf{r})\, v  = \epsilon\, u\; \label{eq:BdGU} \\
\frac{\hbar^2}{2m} \left( \nabla^2 + k_{\bot}^2 \right)\,v
      + \Delta^*(\mathbf{r})\, u  = \epsilon\, v\;. \label{eq:BdGV}
\end{eqnarray}
\end{subequations}
Here $\nabla=\partial_x \mathbf{x}_0+\partial_y \mathbf{y}_0$,
$\mathbf{r}=(x,\;y)$ is a radius vector in the plane perpendicular to the cylinder axis, %
$\Delta(\mathbf{r})$ is the gap function,
$k_{\bot}^2=k_F^2-k_z^2$, $k_z$ is the momentum projection on the
vortex axis.

Following the procedure described in
\cite{Kopnin-PRB07,Melnikov-PRB08,mel-silaev} we introduce the
momentum representation:
\begin{equation}\label{eq:WFPR}
    \hat{\psi}(\mathbf{r}) =
        \left(u \atop v \right) =
        \frac{1}{(2\pi\hbar)^2}\int d^2\mathbf{p}\;%
            \mathrm{e}^{i\mathbf{p}\mathbf{r}/\hbar}\, \hat{\psi}(\mathbf{p})
\end{equation}
where $\mathbf{p} = \vert\mathbf{p}\vert\,%
    (\cos\theta_p\,, \sin\theta_p)= p\, \mathbf{p}_0$.
The unit vector $\mathbf{p}_0$ parametrized by the angle
$\theta_p$ defines the trajectory direction in the ($x,\,y$)
plane. We assume that our solutions correspond to the momentum
absolute values $p$ close to the value $\hbar k_{\bot}$: $p=\hbar
k_{\bot}+q$ ($|q|\ll \hbar k_{\bot}$). Within the quasiclassical
approach the wave function in the momentum representation assumes
the form
%
\begin{equation}\label{eq:FTWFPR}
    \hat\psi(\mathbf{p})= \frac{1}{k_{\bot}}%
        \int\limits_{-\infty}^{+\infty} ds\, %
\mathrm{e}^{i(k_{\bot}-\vert\mathbf{p}\vert/\hbar)\,s}%
            \hat\psi(s,\theta_p)\,.
\end{equation}
Finally, the slowly varying part of the wave function $\hat{\psi}$
in the real space ${\bf r}=r(\cos\theta,\sin\theta)$ is expressed
from Eqs.(\ref{eq:WFPR}, \ref{eq:FTWFPR}) in the following way
(see Ref.~\onlinecite{Kopnin-PRB07}):
\begin{equation}\label{eq:WFCS}
    \hat{\psi}(r,\theta) %
        =\int\limits_{0}^{2\pi} %
            \mathrm{e}^{i k_{\bot} r cos(\theta_p-\theta)} %
            \hat\psi(r \cos(\theta_p-\theta),\theta_p) %
            \frac{d\theta_p}{2\pi}\,,
\end{equation}
where ($r$, $\theta$, $z$) is a cylindrical coordinate system. The
appropriate boundary conditions for wave function $\hat{\psi}$
(\ref{eq:WFPR}) at $y=0$ are follows \cite{BTK-PRB82}:
\begin{subequations}\label{eq:WFBC}
\begin{eqnarray}
    && \hat{\psi}(x,0+) = \hat{\psi}(x,0-) = \hat{\psi}_0\,,   \label{eq:WFBC1} \\%
    && \partial_y \hat{\psi}(x,0+) - \partial_y \hat{\psi}(x,0-) = 2 k_{\bot}
    Z \hat{\psi}_0\,,                                   \label{eq:WFBC2}
\end{eqnarray}
\end{subequations}
where the dimensionless barrier strength $Z = H / \hbar V_{\bot}$
($m V_{\bot} = \hbar k_{\bot}$) defines the transmission $D = 1 /
(1 + Z^2)$ and reflection $Z^2 / (1 + Z^2)$ coefficients in the
normal state.

For extremely weak barrier ($D \gtrsim D_{\xi}$) we can neglect
the anisotropy of the order parameter $\Delta(\mathbf{r})$ around
the vortex and assume that
\begin{equation}\label{eq:OP}
    \Delta(\mathbf{r}) = \Delta_0\, \delta_v(r)\,\mathrm{e}^{i\theta}\,,
        \quad r=\sqrt{x^2+y^2}\,.
\end{equation}
Here $\delta_v(r)$ is a normalized order parameter magnitude for a
vortex centered at $r=0$, such that $\delta_v(r) = 1$ for
$r\to\infty$. Nevertheless the solution (\ref{eq:WFCS}) can not be
characterized by a definite angular momentum $\mu$ because of the
normal reflection of quasiparticles at the defect results in
interaction of angular harmonics with opposite momentum ($\mu$ and
$-\mu$) (see Fig.~\ref{Fig:2}). Thus, following
Ref.~\cite{Kopnin-PRB03} we introduce the angular momentum
expansion for the solution (\ref{eq:WFCS}):
\begin{equation}\label{eq:WFAEXP}
    \hat{\psi}(s,\,\theta_p) = \sum\limits_{\mu} \mathrm{e}^{i \mu\theta_p + i\,\hat{\sigma}_z \theta_p /\, 2}
        \hat{f}{_\mu}(s)\,,
\end{equation}
where $\mu = n + 1/2$, and $n$ is an integer. The function
$\hat{f}{_\mu}(s)$ satisfies the Andreev equation along the
quasiclassical trajectory with the impact parameter $b = - \mu /
k_{\bot}$
\begin{equation}\label{eq:QCE0}
    -i \hbar V_{\bot} \hat{\sigma}_z\,\partial_s\hat{f}_{\mu}
              + \hat{\Delta}_b(s) \hat{f}_{\mu} = \varepsilon \hat{f}_{\mu}\,,
\end{equation}
where
\begin{equation}\label{eq:QCE0-D}
    \hat{\Delta}_b(s) = \hat{\sigma}_x\,\mathrm{Re}D_b(s) %
                       -\hat{\sigma}_y\,\mathrm{Im}D_b(s)
\end{equation}
is the gap operator, and $\hat{\sigma}_i$ are the Pauli matrices.
Taking into account the evident relations
\begin{eqnarray}
x = s\cos\theta_p- b\sin\theta_p\,, \quad
y = s\sin\theta_p + b\cos\theta_p\,,       \nonumber \\
x \pm i y = (s \pm i b)\,%
    \mathrm{e}^{\pm i \theta_p}\,.\qquad\qquad  \nonumber
\end{eqnarray}
one obtains from (\ref{eq:OP}) the following expression for the
order parameter $\Delta(\mathbf{r})$ around the vortex in ($s$,
$\theta_p$) variables:
\begin{equation}\label{eq:OPST}
    \Delta = D_b(s)\,\mathrm{e}^{i\theta_p}\,,\quad
             D_b(s)=\Delta_0\,\frac{\delta_v(\sqrt{s^2+b^2})}
                                   {\sqrt{s^2+b^2}}(s+i b)\,.
\end{equation}
Changing the sign of the coordinate $s$ one can observe a useful
symmetry property of the solution of Eq.(\ref{eq:QCE0}):
\begin{equation}\label{eq:QCE0-SYM}
\hat f_{\mu} (-s) = \pm\,\hat{\sigma}_y \hat f_{\mu} (s) \ .
\end{equation}


\subsection{General solution}\label{GenSol}

To find the solution of Eqs.~(\ref{eq:QCE0},\ref{eq:QCE0-D}) we
can use the results of Ref.~\onlinecite{Kopnin-PRB07}. For low
energies ( $\mu \ll k_{\bot} \xi$ ) we take the function
$\hat{f}_{\mu}$ as a sum
\begin{equation}\label{eq:GESO0}
    \hat f_{\mu} = c_{\mu 1}  \hat{G}_{\mu 1} + c_{\mu 2} \hat{G}_{\mu 2}
\end{equation}
of the two linearly independent solutions
\begin{subequations}\label{eq:GESO}
\begin{eqnarray}
   &&\hat{G}_{\mu 1} =   \mathrm{e}^{i\,\hat{\sigma}_z \pi/\, 4} \left(
        \mathrm{e}^{-|D(s)|} - i\, \mathrm{sgn}(s) \frac{\gamma_{\mu}}{2} \hat{\sigma}_z
        \mathrm{e}^{|D(s)|} \right) \hat{\lambda} \,,  \qquad               \label{eq:GESO1} \\
   &&\hat{G}_{\mu 2} =   \mathrm{e}^{i\,\hat{\sigma}_z \pi/\, 4}
        \mathrm{e}^{-|D(s)|}\, \hat{\sigma}_z\, \hat{\lambda} \,,           \label{eq:GESO2}
\end{eqnarray}
\end{subequations}
where  $\hat{\lambda}=\left( 1 , 1 \right)^T$\,,
\begin{eqnarray}
   &&D(s) = \frac{k_F}{k_{\bot} \xi}\,\int\limits_0^{s} dt\,%
            \frac{ t\, \delta_v \left(\sqrt{t^{2}+b^{2}}\right)}
             {\sqrt{t^{2}+b^{2}}}\,,                                        \label{eq:GESO3} \\
   &&\Lambda_{\mu} = \frac{2\, k_F}{k_{\bot} \xi}\,
        \int\limits_{0}^{\infty} ds\,\mathrm{e}^{-2 D(s)}\,,                 \label{eq:GESO4} \\
   &&\gamma_{\mu} = \frac{\Lambda_{\mu}}{\Delta_0}
        \left(\, \varepsilon_{\mu} - \varepsilon\, \right)                   \label{eq:GESO5}
\end{eqnarray}
and
\begin{equation}\label{eq:CdGM}
    \varepsilon_{\mu} =
        -\frac{ 2 \Delta_0\, k_F \mu}{k_{\bot}^2 \xi\, \Lambda_{\mu}} \int\limits_{0}^{\infty} ds\,
        \frac{ \delta_v \left(\sqrt{s^{2}+b^{2}}\right)}
             {\sqrt{s^{2}+b^{2}}}\,
             \mathrm{e}^{-2 D(s)}\,
\end{equation}
is the CdGM excitation spectrum. Here $\xi = \hbar V_F / \Delta_0$
is the coherence length ($V_F$ is the Fermi velocity).


\subsection{Boundary condition.}\label{BounCond}

As a next step we rewrite the boundary condition (\ref{eq:WFBC})
for wave functions $\hat{f}_{\pm \mu}(s)$ defined at the
trajectories $s_1$ and $s_2$ (see Fig.~\ref{Fig:1}). Due to normal
reflection of quasiparticles at the defect the trajectories $s_1$
and $s_2$ with opposite momentum ($\mu$ and $-\mu$) directions are
coupled. Substituting the expressions
(\ref{eq:WFCS},\ref{eq:WFAEXP}) into the boundary condition
(\ref{eq:WFBC}),
we obtain the
following relation between the amplitudes of incident
$\hat{f}_{\pm\mu i}(s)$ and transmitted $\hat{f}_{\pm\mu t}(s)$
two-component quasiparticle wave functions at the point $s_0 = - b
/ \tan\theta_p$ where the trajectories cross the barrier: 
\begin{equation}\label{eq:WFBC-FMU}
   (\eta + i)\, \hat{f}_{\pm\mu t} = \eta\, \hat{f}_{\pm\mu i} -
        i \mathrm{e}^{\mp i\,\hat{\sigma}_z \theta_p} \hat{f}_{\mp \mu i}\,,
\end{equation}
where $\eta = \sin\theta_p / Z$. Our further analysis of
quasiparticle excitations is based on the solutions
(\ref{eq:GESO0},\ref{eq:GESO}) which must be supplemented by the
boundary conditions (\ref{eq:WFBC-FMU}).


\section{Spectrum of the vortex pinned
by planar defect}\label{SingleQuant}

We now proceed with the analysis of the subgap spectrum for a
singly quantized vortex trapped by the planar defect. Hereafter in
this section we assume the angular momentum to be positive, i.e.
$\mu >0$. The form of the two-component quasiparticle wave
functions $\hat{f}_{\pm\mu}(s)$ depends on a position of the point
$s_0$ at the trajectory. If the coordinate $s_0 \ge 0$ than the
general solution (\ref{eq:GESO0},\ref{eq:GESO}) takes the
following form
\begin{equation}
    \hat{f}_{\pm\mu}(s) = \left\{
            \begin{array}{c}
                c_{\pm\mu i}\, \mathrm{e}^{i (\hat{\sigma_z} \mp 1) \pi /4}
                \mathrm{e}^{-|D(s)|}\, \hat{\lambda} \,,
                \\  s \le 0 \,, \\
               c_{\pm\mu i}\, \mathrm{e}^{i (\hat{\sigma_z} \mp 1) \pi /4} \left(
               \mathrm{e}^{-|D(s)|} - i \gamma_{\pm\mu} \hat{\sigma_z} \mathrm{e}^{|D(s)|} \right)\, \hat{\lambda} \,,
               \\  0 \le s \le s_0 \,, \\
                c_{\pm\mu t}\,\mathrm{e}^{i (\hat{\sigma_z} \mp 1) \pi /4}  \mathrm{e}^{-|D(s)|}\, \hat{\lambda} \,,
                \\ s \ge s_0  \,,
            \end{array}
                \right. \,,  \label{eq:WF-FMU1}
\end{equation}
where
$$
    \gamma_{+\mu} = -\frac{\Lambda_{\mu}}{\Delta_0} \left( | \varepsilon_{\mu}
    | + \varepsilon \right)\,, \quad
    \gamma_{-\mu} = \frac{\Lambda_{\mu}}{\Delta_0} \left( | \varepsilon_{\mu}
    | - \varepsilon \right)\,.
$$
 Otherwise, if $s_0 \le 0$
\begin{equation}
    \hat{f}_{\pm\mu}(s) = \left\{
            \begin{array}{c}
                c_{\pm\mu i}\, \mathrm{e}^{i (\hat{\sigma_z} \mp 1) \pi /4}
                \mathrm{e}^{-|D(s)|}\, \hat{\lambda} \,,
                \\  s \le s_0  \,, \\
               c_{\pm\mu t}\, \mathrm{e}^{i (\hat{\sigma_z} \mp 1) \pi /4} \left(
               \mathrm{e}^{-|D(s)|} + i \gamma_{\pm\mu} \hat{\sigma_z} \mathrm{e}^{|D(s)|} \right)\, \hat{\lambda} \,,
               \\  s_0 \le s \le 0 \,, \\
                c_{\pm\mu t}\, \mathrm{e}^{i (\hat{\sigma_z} \mp 1) \pi /4}
                \mathrm{e}^{-|D(s)|}\, \hat{\lambda} \,,
                \\ s \ge 0  \,,
            \end{array}
                \right. \label{eq:WF-FMU2}
\end{equation}
The eigenfunction $\hat{f}_{\pm\mu}(s)$ has to be normalized
$$
    \int\limits_{-\infty}^{\infty} d s \left( | \hat{f}_{+\mu}(s) |^2
    + | \hat{f}_{-\mu}(s) |^2 \right) = k_{\bot} \,.
$$

Substituting the above expressions (\ref{eq:WF-FMU1}) or
(\ref{eq:WF-FMU2}) into the boundary conditions
(\ref{eq:WFBC-FMU}), we obtain the following system of algebraic
equations with respect to the amplitude $ c_{\pm\mu i}$ of the
incident waves
\begin{subequations}\label{eq:AM-FMU}
\begin{eqnarray}
   \eta\,\gamma_{+\mu}\,c_{+\mu i} + \left( \gamma_{\mp\mu}\,\cos\theta_p
        + \mathrm{e}^{-2 D_0}\,\sin\theta_p \right)\,c_{-\mu i} = 0 \,, \quad   \label{eq:AM-FMU1} \\
   \eta\,\gamma_{-\mu}\,c_{-\mu i} - \left( \gamma_{\pm\mu}\,\cos\theta_p
        - \mathrm{e}^{-2 D_0}\,\sin\theta_p \right)\,c_{+\mu i} = 0 \,. \quad   \label{eq:AM-FMU2}
\end{eqnarray}
\end{subequations}
The case $s_0 \ge 0$ ($s_0 < 0$) corresponds to the choice of
upper (lower) sign in Eqs.~(\ref{eq:AM-FMU}), $D_0 = D(s_0)$ and
the angle $\theta_p$ defines the direction of the ray with the
angular momentum $+\mu$. To find the subgap quasiparticle
excitation spectrum we should find the determinant of the
algebraic system, and its zero give us the equation for the energy
spectrum $\varepsilon$:
\begin{eqnarray}
   \varepsilon^2(b\,,\theta_p)&=& \varepsilon_{\mu}^2 + \left( \frac{\Delta_0}{\Lambda_{\mu}} \right)^2
        \frac{\mathrm{e}^{-2 D_0}}{\eta^2 + \cos^2\theta_p} \times       \label{eq:SPECTRUM} \\
       &&\left[ \Lambda_{\mu} \frac{|\varepsilon_{\mu}|}{\Delta_0} |\sin(2 \theta_p)|
            + \mathrm{e}^{-2 D_0}\, \sin^2\theta_p \right]\,.
            \nonumber
\end{eqnarray}

\begin{figure}
\includegraphics[width=0.40\textwidth]{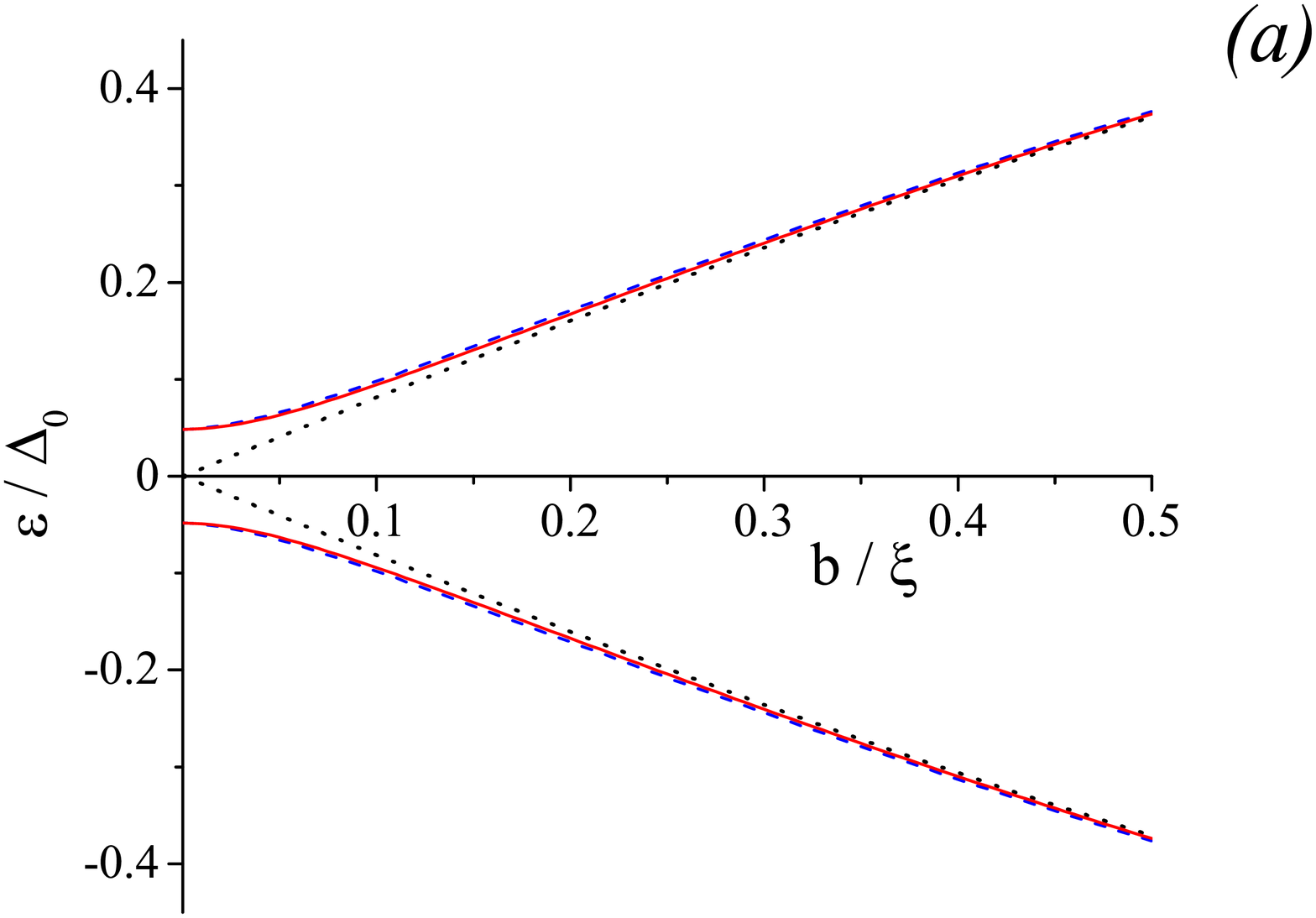}
\includegraphics[width=0.40\textwidth]{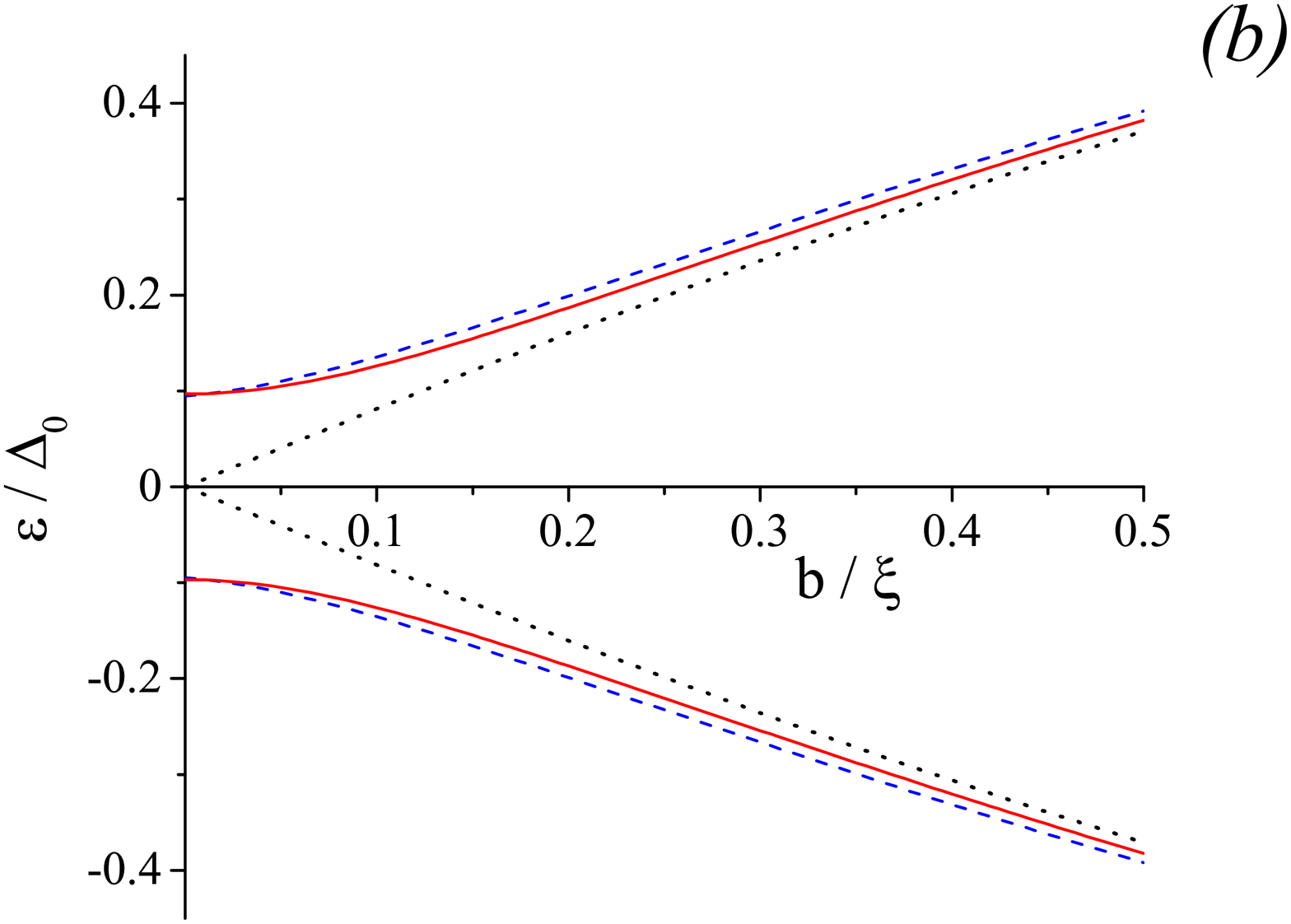}
\includegraphics[width=0.40\textwidth]{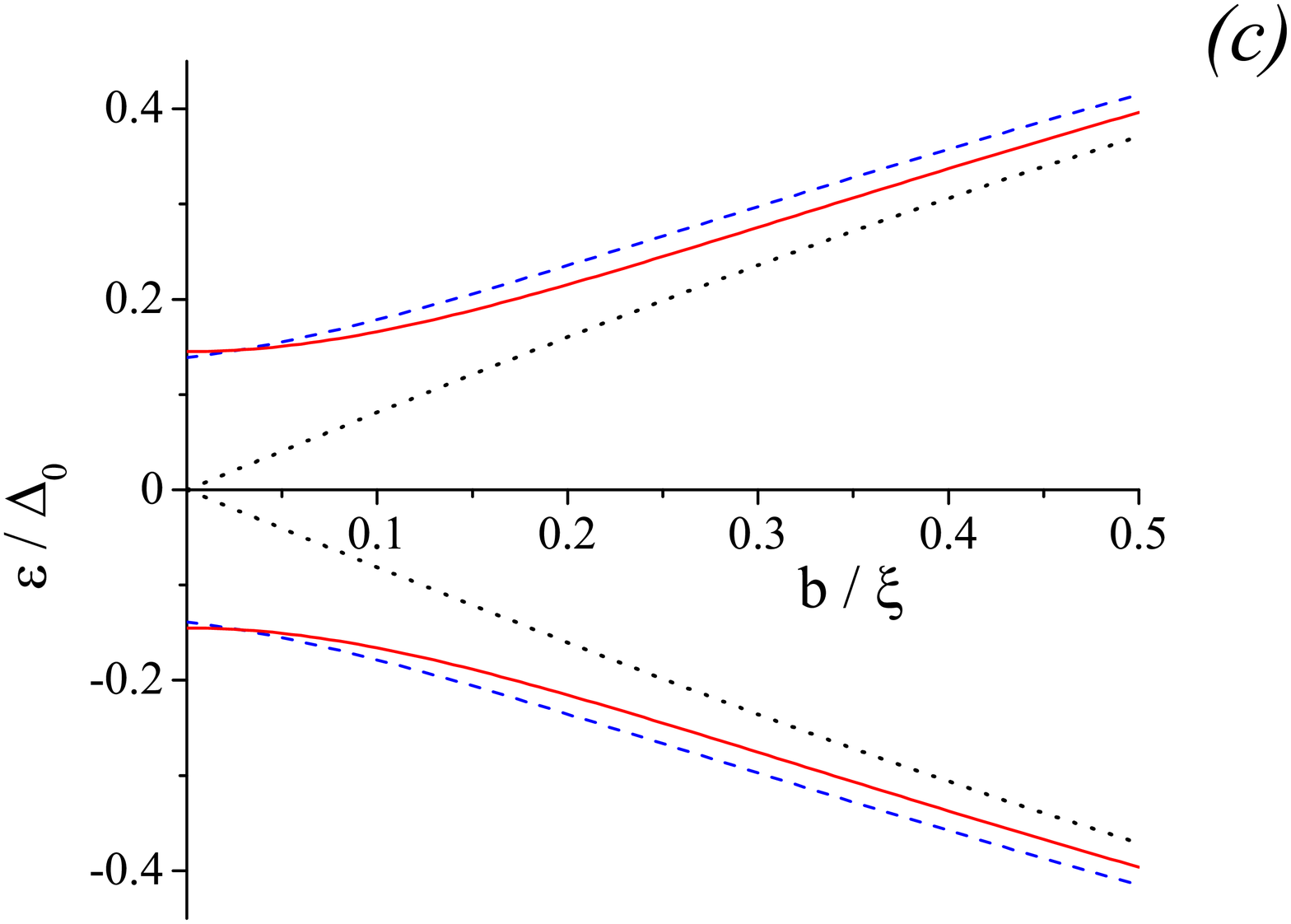}
\caption{(Color online) Quasiparticle spectra $\varepsilon(b,\,
\theta_p)$ calculated using Eq.~(\ref{eq:SPECTRUM}) for different
values of the dimensionless barrier strength $Z$ and the
trajectory direction $\theta_p$ in the ($x,\,y$) plane ($k_z =
0$): ($a$)  $Z = 0.1$; ($b$)  $Z = 0.2$; ($c$)  $Z = 0.3$. Dotted
lines for $\theta_p = 0$ correspond to the CdGM branch of the
spectrum. The dash blue lines show the dependence for $\theta_p =
\pi /4$; solid red lines show the dependence for $\theta_p = \pi
/2$.} \label{Fig:4}
\end{figure}

Figure~\ref{Fig:4} shows the anomalous spectral branches as
functions of the impact parameter $b = -\mu / k_F$ for different
values of the dimensionless barrier strength $Z$ and the
trajectory directions in the ($x,\,y$) plane determined by the
angle $\theta_p$. The qualitative behavior of the spectrum is
weakly sensitive to the concrete profile of the gap amplitude
inside the core and we choose a simple model dependence
\begin{equation}\label{eq:GAP}
    \delta_v(r) = r / \sqrt{r^2 + \xi^2}
\end{equation}
neglecting, thus, the influence of the defect on the behavior of
the gap profile.
%
%
Contrary to the CdGM case the spectrum branch (\ref{eq:SPECTRUM})
does not cross the Fermi level in presence of the defect. For
rather small $Z$ the minigap in the quasiparticle spectrum
$$
\Delta_{m}(\theta_p)=\varepsilon(0, \theta_p) = \frac{\Delta_0
}{\Lambda_0} \frac{Z}{\sqrt{ 1 + Z^2 / \tan^2 \theta_p}}
$$
appears to be almost independent of $\theta_p$ in a wide range of angles except
the small angular intervals close to $\theta_p=0$ and $\theta_p=\pi$.
It is natural to expect that in the patterns of the local density of states (LDOS)
this angular independent quantity
 should reveal itself as a soft gap $\Delta_{Soft}\sim Z\Delta_0$ growing with
the increasing barrier strength $Z$ (see the
Section~\ref{LocCond}). We emphasize here the fact that this gap
is soft since the spectrum  (\ref{eq:SPECTRUM}) for small
$|\tan\theta_p|\lesssim Z$ is gapless and, thus, these angular
intervals can contribute to the LDOS at the Fermi level. This
nonzero contribution exists, of course, only in the quasiclassical
limit when we completely neglect the quantum mechanical nature of
the trajectory precession which should be responsible for the
opening of the hard minigap for the energies below
$\Delta_{Soft}$.

\begin{figure}
\includegraphics[width=0.45\textwidth]{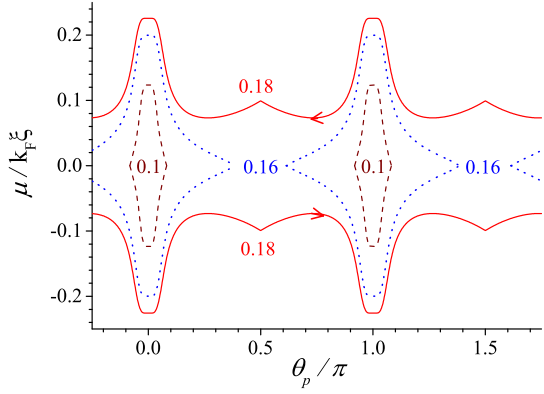}
\caption{(Color online) Quasiparticle orbits in the
$\mu\,-\,\theta_p$ plane corresponding to different energy levels
for the dimensionless barrier strength $Z = 0.3$. The numbers near
the curves denote the corresponding values of $\varepsilon /
\Delta_0$. The direction of trajectory precession along the orbits
is shown by arrow. We put here $k_z = 0$.} \label{Fig:5}
\end{figure}

To derive the corresponding quantization rules in the limit $Z \ll
1$ we consider isoenergetic lines $\mu(\theta_p) = - k_{\bot}
b(\theta_p)$ in ($\mu\, -\, \theta_b$) plane. The resulting
classical orbits are shown in Fig.~\ref{Fig:5}. Generally, one can
distinguish two types of the isoenergetic lines behavior. If the
quasiparticle energy is of the order of the minigap ($\varepsilon
\lesssim \Delta_{Soft}$) there appear prohibited angular domains
centered at the points $\theta_p = \pm \pi/2$ due to the normal
reflection of quasiparticles at the defect. In this case classical
orbits form close paths in ($\mu\,-\,\theta_b$) space
corresponding to the precession of the trajectory in the region
with the width $2\, \delta \theta_p(\varepsilon)$ near the points
$\theta_p = 0,\, \pm \pi$. The width $2\, \delta\theta_p$ of the
precession region grows with an increase in energy level. For
small $| \mu | \ll k_{\bot} \xi$ the value $\delta\theta_p$ can be
estimated as follows:
\begin{equation}\label{eq:WIDTH}
    \delta\theta_p \simeq \frac{\varepsilon\,\Lambda_0 / \Delta_0}{\sqrt{ 1 - (\varepsilon\,\Lambda_0 /
    Z \Delta_0)^2}}.
\end{equation}
Shrinking of the prohibited angular domains and the crossover from the closed
orbits to the open ones occur at the energy $\varepsilon^*$
satisfying the condition $\delta\theta_p(\varepsilon^*) = \pi/2$. 
%
%

\begin{figure}
\includegraphics[width=0.45\textwidth]{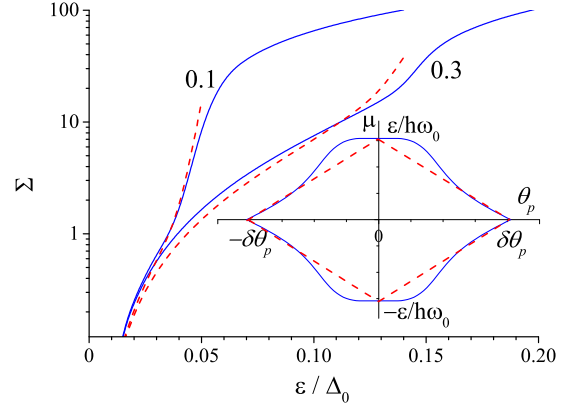}
\caption{(Color online) Dependence $\Sigma(\varepsilon)$
(\ref{eq:Bohr-Somm}) for two values of the dimensionless barrier
strength $Z$. Results of numerical calculations are shown by the
blue solid lines. Dashed red curves show approximate values of
$\Sigma(\varepsilon)$ obtained from Eq.~(\ref{eq:Bohr-Somm-Ap}).
The insert shows quasiparticle orbit in the ($\mu\,-\,\theta_b$)
plane (blue solid line) and its approximation (red dashed line)
described by the equation
 (\ref{eq:Bohr-Somm-Ap}). We put here $k_z = 0$ and
$E_F/\Delta_0=\mathrm{50}$. } \label{Fig:6}
\end{figure}

The low lying energy levels of quasiparticles can be obtained by
applying the Bohr-Sommerfeld quantization rule
(\ref{eq:Bohr-Somm}) for closed paths in the plane of canonically
conjugate variables $\mu$ and $\theta_p$. Figure~\ref{Fig:6} shows
the typical dependence $\Sigma(\varepsilon)$ calculated using the
spectrum (\ref{eq:SPECTRUM}). Taking $\varepsilon_{\mu} \simeq -
\hbar \omega_0 \mu$ for small $\mu$ values and replacing the real
classical orbits in ($\mu-\theta_b$) plane by the model one (see
the insert Fig.~\ref{Fig:6}), one can obtain a reasonable fit
(dashed curve) to the numerical results (solid curve):
\begin{equation}\label{eq:Bohr-Somm-Ap}
    \Sigma(\varepsilon) \approx 2 \frac{\varepsilon}{\hbar \omega_0}
    \delta\theta_p = \frac{2 \varepsilon^2 \Lambda_0 / \Delta_0}{\hbar \omega_0\,
    \sqrt{ 1 - (\varepsilon\,\Lambda_0 / Z \Delta_0)^2}}.
\end{equation}
The above relation together with the Bohr-Sommerfeld rule
(\ref{eq:Bohr-Somm}) results in the following explicit expression
for discrete subgap energy levels
\begin{eqnarray}
    \varepsilon_n &\simeq& \frac{\Delta_0 Z}{\Lambda_0}
    \left[\, p_n \sqrt{1 + p_n^2 /4} - p_n^2 / 2\,
    \right]^{1/2}\,,\label{eq:Spacing} \\
    &&p_n = \frac{\pi \Lambda_0 \Delta_0}{2 E_F Z^2}\,
    ( n + \beta )\,,   \nonumber
\end{eqnarray}
which appears to be justified for $\varepsilon_n / \Delta_0
\lesssim Z^2 \ll 1$. The expression (\ref{eq:Spacing}) can be
strongly simplified provided $p_n \ll 1$ for low lying energy
levels:
$$
    \varepsilon_n^2 \simeq \frac{\pi}{2 \Lambda_0}
        \frac{\Delta_0^3}{E_F}\, ( n + \beta )
        \left[\,1 - \frac{\pi \Lambda_0 \Delta_0}{4 E_F Z^2}\, ( n + \beta )\, \right]\,.
$$
The main term of the last relation appears to be in good agreement
with the estimate (\ref{eq:Spacing0}) and describes qualitatively
the new behavior of spectrum of subgap quasiparticle states
($\varepsilon_n \sim n^{1/2}$) due to the normal scattering at the
planar defect. Both the hard minigap $\varepsilon_0 \lesssim
\Delta_0 \sqrt{\Delta_0 / E_F} \ll \Delta_{Soft}$ in the discrete
spectrum (\ref{eq:Spacing}) and the interlevel spacing $\hbar
\omega = \varepsilon_n - \varepsilon_{n-1}$ grow with the increase
in the barrier strength $Z$.
%
%

Besides its fundamental interest, the problem of pinned vortex
spectrum important for understanding the nature of dissipation in
the presence of planar defects. In particular, according to the
spectral flow theory \cite{Kopnin-TNSC}, it is the behavior of the
anomalous branch which determines the high-frequency conductivity
and Kerr effect
\cite{Rosenstein-PRB11_microwave-absorp,Vadimov-Melnikov-JLTP16_chiral}.
One can expect that the opening of the hard minigap
$\varepsilon_0$ in discrete quasiparticle spectrum
(\ref{eq:Spacing}) and change in the slope $\varepsilon(\mu)$
dependence (\ref{eq:SPECTRUM}) can cause the suppression of the
dissipation accompanying the vortex motion and the resulting
changes in the relation between the Ohmic and Hall conductivities.
As a result, the quasiparticle subgap spectrum can be tested by
the measurements of the conductivity tensor at finite frequencies.

%

\section{Local density of states
for a  pinned vortex}\label{LocCond}

We now proceed with the calculations of the local density of
states for a singly quantized vortex pinned at the planar defect.
This quantity is known to be directly probed in the scanning
tunneling microscopy/spectroscopy experiments
\cite{Roditchev-NatPh15}. For the sake of simplicity we assume
here the Fermi surface to be a cylinder and neglect the dependence
of the quasiparticle energy on the momentum component $k_z$ along
the cylinder axis $z$ considering a motion of quasiparticles only
in ($x,\,y$) plane.
%
\begin{figure}[t!]
\includegraphics[width=0.45\textwidth]{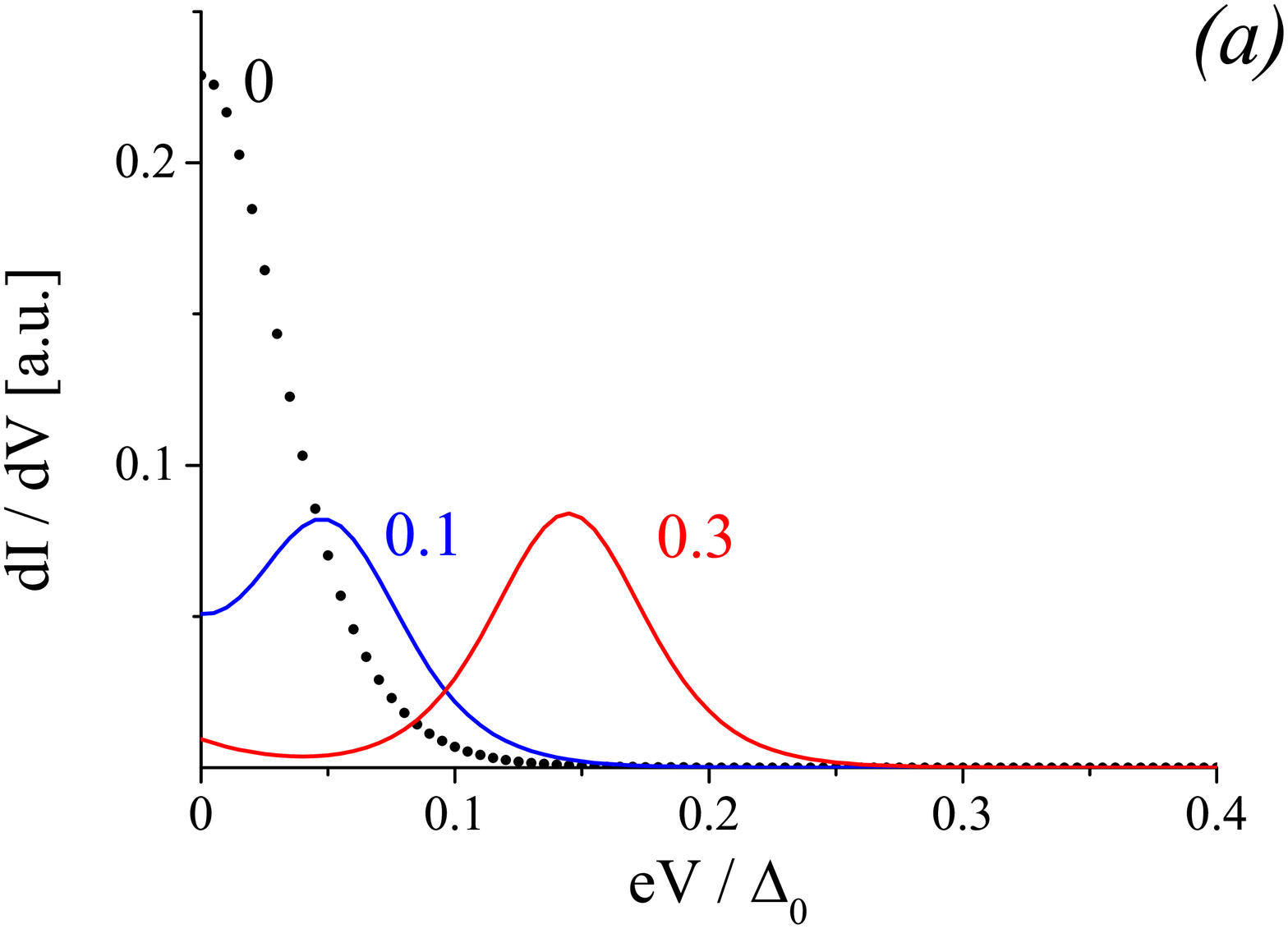}
\includegraphics[width=0.45\textwidth]{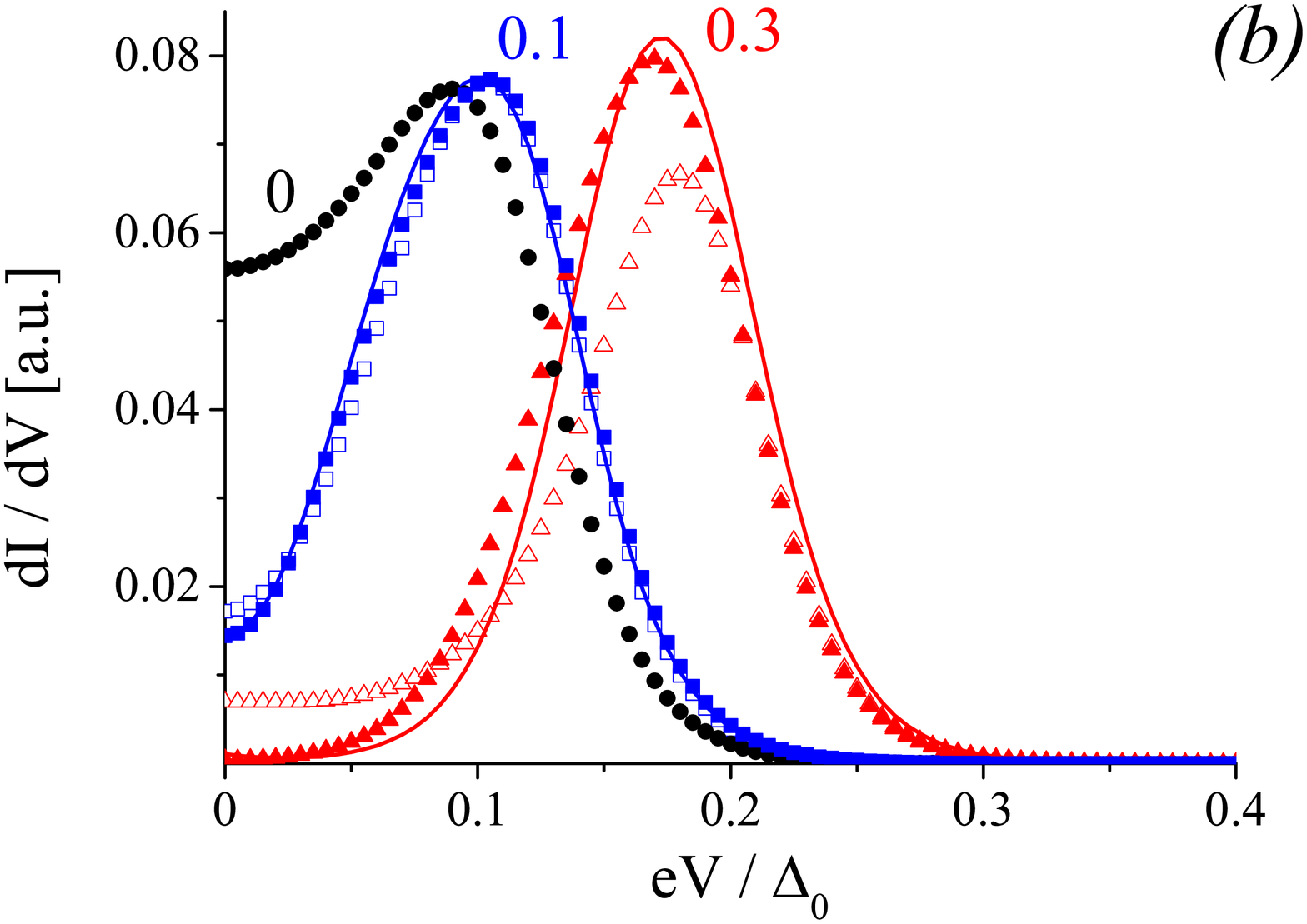}
\includegraphics[width=0.45\textwidth]{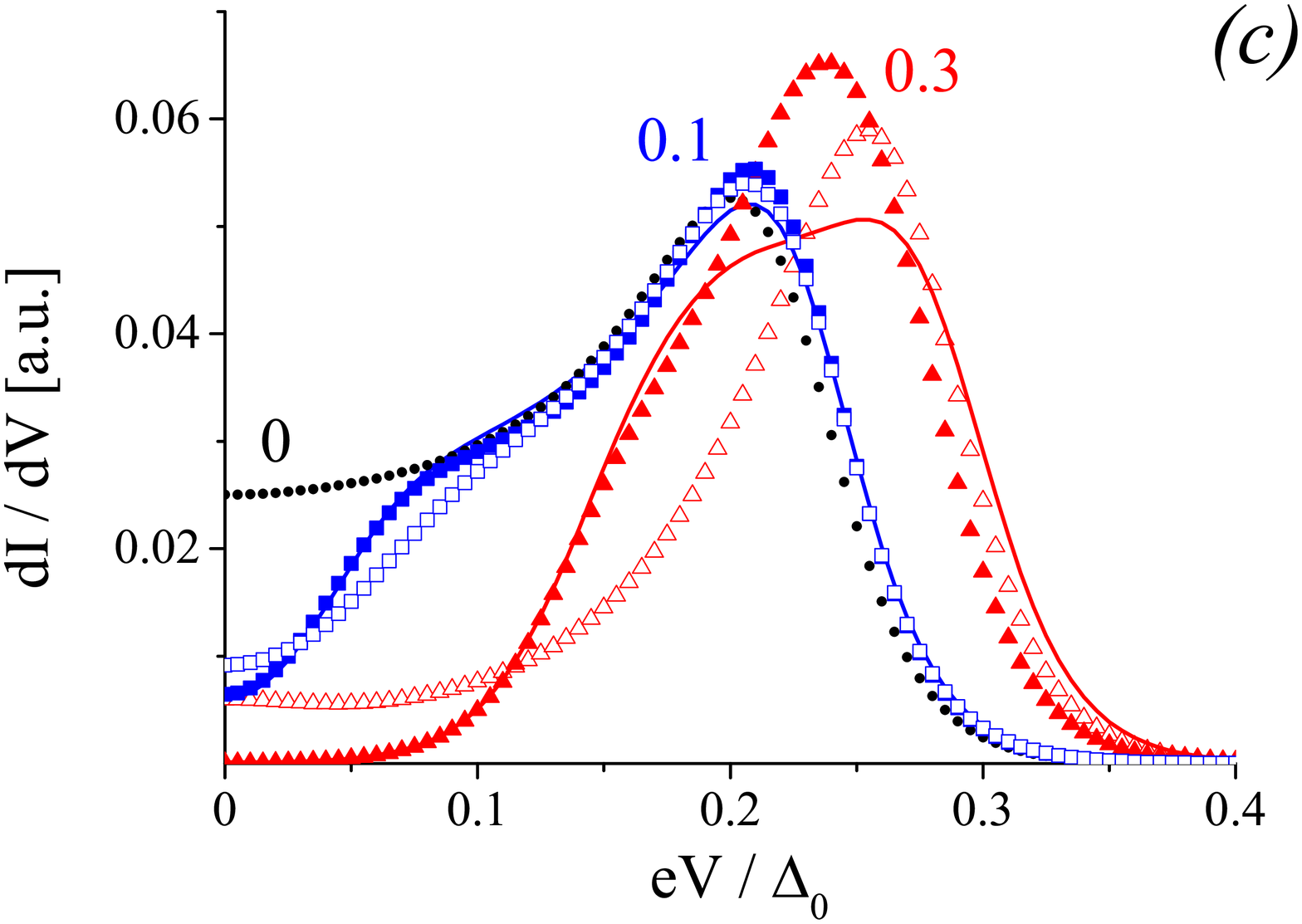}
\caption{(Color online) Local differential conductance $dI/dV$
versus bias voltage ($eV$) in different points ($r,\,\theta$) at
the plane ($x,\,y$): ($a$) $r=0$; ($b$) $r = 0.14\, \xi$; ($c$) $r
= 0.28\, \xi$. The numbers near the curves denote the
corresponding values of the dimensionless barrier strength $Z$.
The lines correspond to the case $\theta = \pi/4$; empty symbols
-- $\theta = 0$; filled symbols -- $\theta = \pi / 2$. We put here
$T/\Delta_0=\mathrm{0.02}$. For reference black filled circles
($\bullet$) show the local $dI/dV$ curves for the free Abrikosov
vortex ($Z=0$).} \label{Fig:7}
\end{figure}
%
The peculiarities of the LDOS are usually determined from the
analysis of the local differential conductance (LDC):
\begin{equation}\label{eq:LDC1}
    \frac{d I/d V}{\left(d I/d V\right)_N} = %
        \int\limits_{-\infty}^{\infty} d\varepsilon\, %
        \frac{N(\mathbf{r},\,\varepsilon)}{N_0}
        \frac{\partial f(\varepsilon-eV)}{\partial V}\,,
\end{equation}
where $V$ is the applied voltage, $(dI/dV)_N$ is a conductance of
the normal metal junction, and $f(\varepsilon) = 1/\left(1 +
\exp(\varepsilon/T )\right)$ is a Fermi function. Within the
quasiclassical approach the LDOS
\begin{equation}\label{eq:LDC2}
    N(\mathbf{r},\,\varepsilon) =
        k_F\,\int d b\, {\vert u_{b}(\mathbf{r}) \vert}^2\,
        \delta(\varepsilon-\varepsilon(b))
\end{equation}
can be expressed through the electron component $u(r,\,\theta)$ of
quasiparticle eigenfunctions (\ref{eq:WFPR}) corresponding to the
energy $\varepsilon(b,\,\theta_p)$ determined by
Eqs.~(\ref{eq:SPECTRUM}),(\ref{eq:GESO3}),(\ref{eq:GESO4}),(\ref{eq:GESO5}),(\ref{eq:CdGM}).
The wave function $\hat{\psi}(r,\theta)$ parametrized by the
impact parameter $b = - \mu / k_F$
\begin{eqnarray}
    &&\hat{\psi}(r,\theta) =
     \left( u(r,\theta) \atop v(r,\theta) \right) =                 \label{eq:LDC3} \\
    &&\quad \mathrm{e}^{ i(2\mu+\hat{\sigma}_z)\theta/2 }
      \int\limits_{0}^{2\pi} \frac{d\alpha}{2\pi}\, %
      \mathrm{e}^{i k_F r \cos\alpha + i(2\mu+\hat{\sigma}_z)\alpha/2}\,
      \hat{f}_{\mu}(r \cos\alpha)\, \nonumber
\end{eqnarray}
in the limit $k_F\, r \gg 1$ can be evaluated using the stationary
phase method. For an impact parameter $\vert b \vert \le r$ the
stationary phase points are given by the condition:
$\sin\alpha_{1,2}= - b/r$. Summing over two contributions in the
vicinity of the stationary angles $\alpha_1 = \theta_{p1}-\theta =
\alpha_r$ and $\alpha_2 = \theta_{p2}-\theta = \pi-\alpha_r$, we
can write the electron component $u(r,\theta)$ of quasiparticle
eigenfunctions as follows:
\begin{eqnarray}
    &&u(r,\,\theta) = \left(\frac{1}{2\pi k_F s_r}\right)^{1/2}
        \mathrm{e}^{i(2\mu+1)\theta/2} \label{eq:LDC4}\\
    &&\quad \times\left[\,f_{\mu}^u(s_r)\,\mathrm{e}^{i\varphi_r}
       + f_{\mu}^u(-s_r)\,\mathrm{e}^{-i\varphi_r+i(2\mu+1)\pi/2}\, \right]\,,\nonumber
\end{eqnarray}
where $s_r=r\vert \cos\alpha_r \vert=\sqrt{r^2-b^2}$.
The phase
$$
    \varphi_r=k_F r \cos\alpha_r + |\mu|\alpha_r
            + \mathrm{sgn}(\mu)\, \alpha_r/2 - \pi/4
$$
is determined by the trajectory orientation angle
$\alpha_r=-\arcsin(b/r)$. Neglecting  the oscillations at the
atomic length scale we obtain the following slowly varying
envelope function:
\begin{equation} \label{eq:LDC5}
    {\vert u(r,\,\theta) \vert}^2 \simeq
        \frac{1}{2 \pi k_F s_r}\, \left[\, {\vert f_{\mu}^u(s_r) \vert}^2
        + {\vert f_{\mu}^u(-s_r) \vert}^2\, \right]\,,
\end{equation}
where the function $f_{\mu}^u(\pm s_r)$ is determined by the relations
(\ref{eq:WF-FMU1}) or (\ref{eq:WF-FMU2}).
%
\begin{figure*}[t!]
\includegraphics[width=0.4\textwidth]{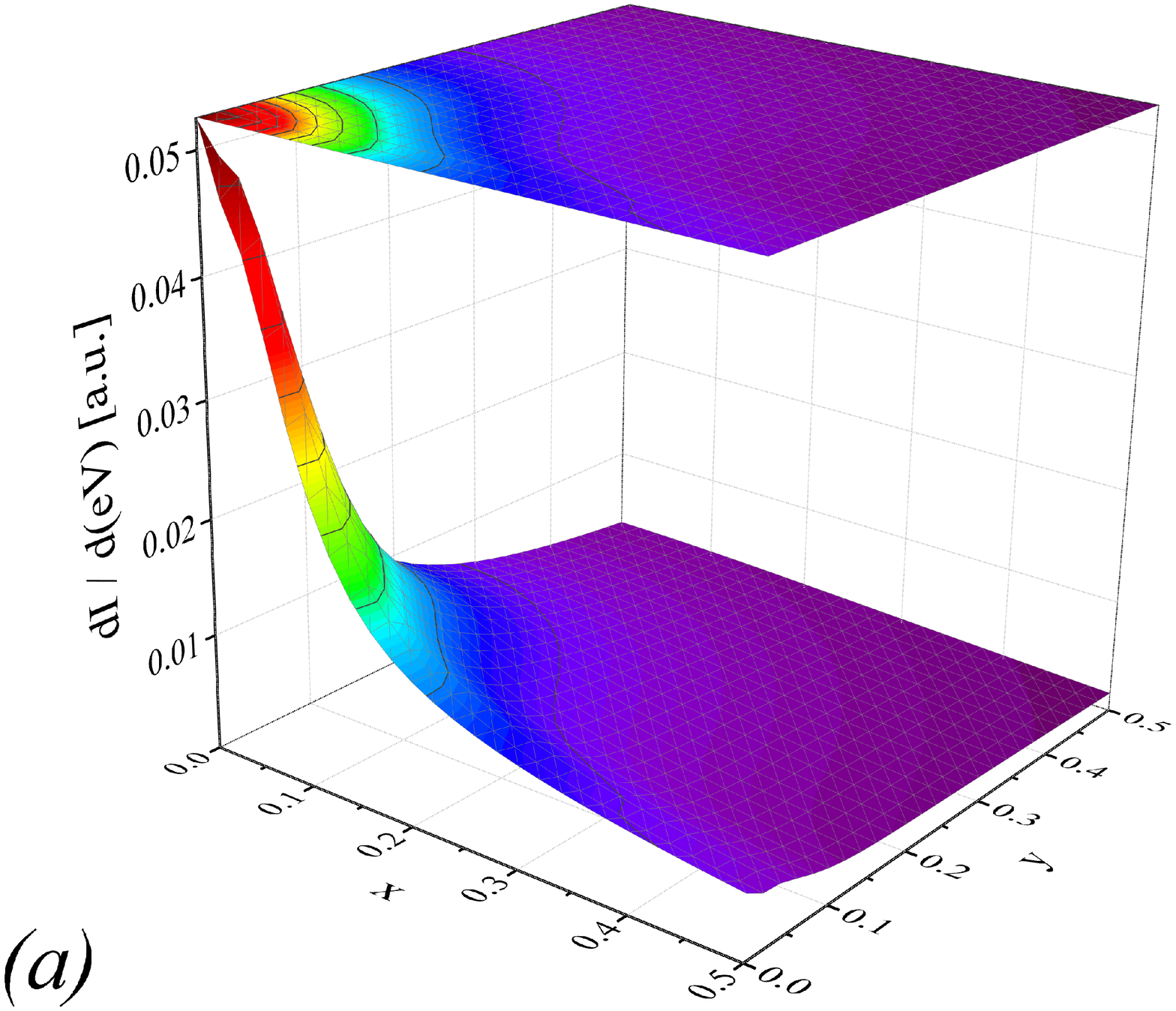}
\includegraphics[width=0.4\textwidth]{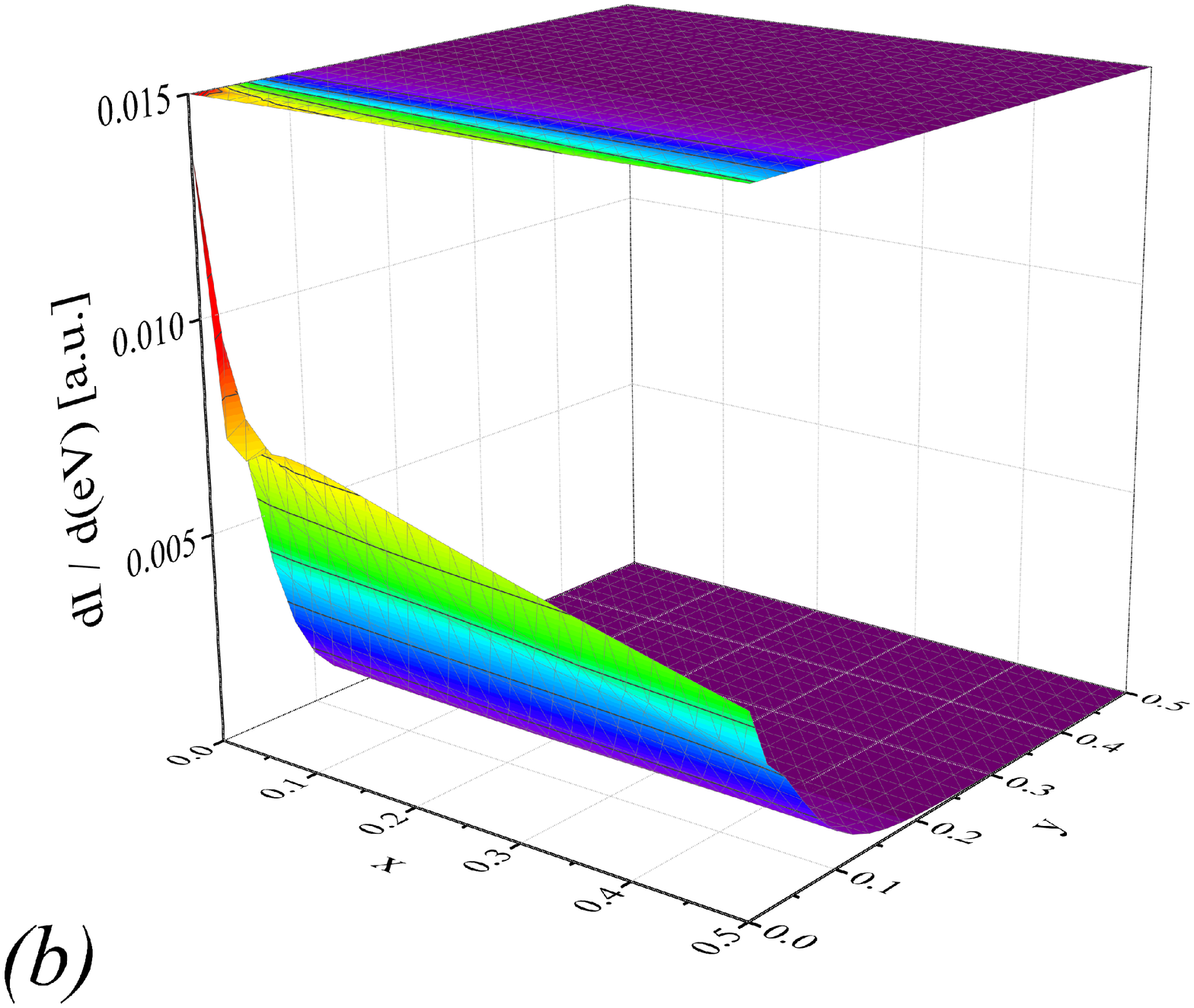}
\includegraphics[width=0.4\textwidth]{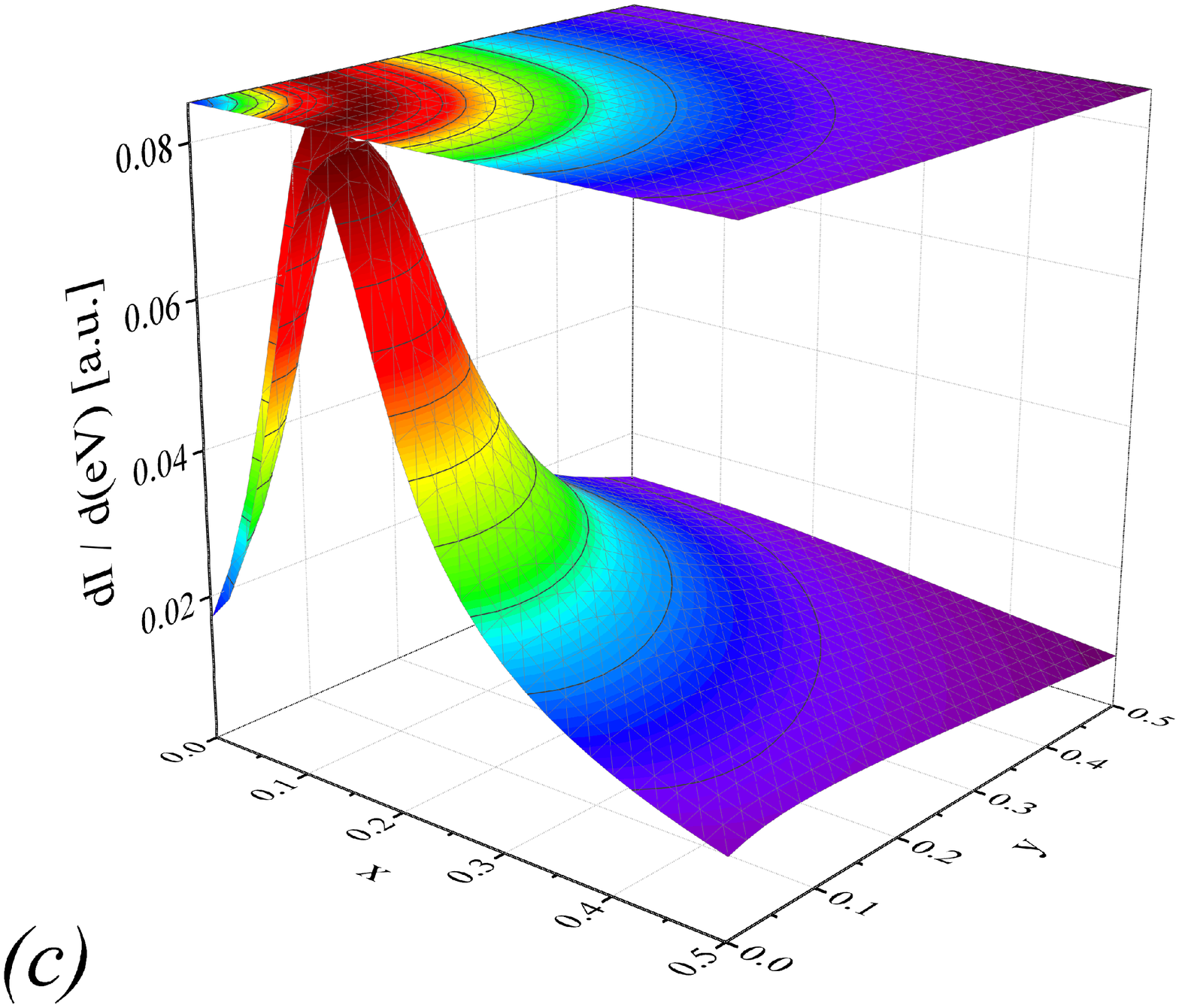}
\includegraphics[width=0.4\textwidth]{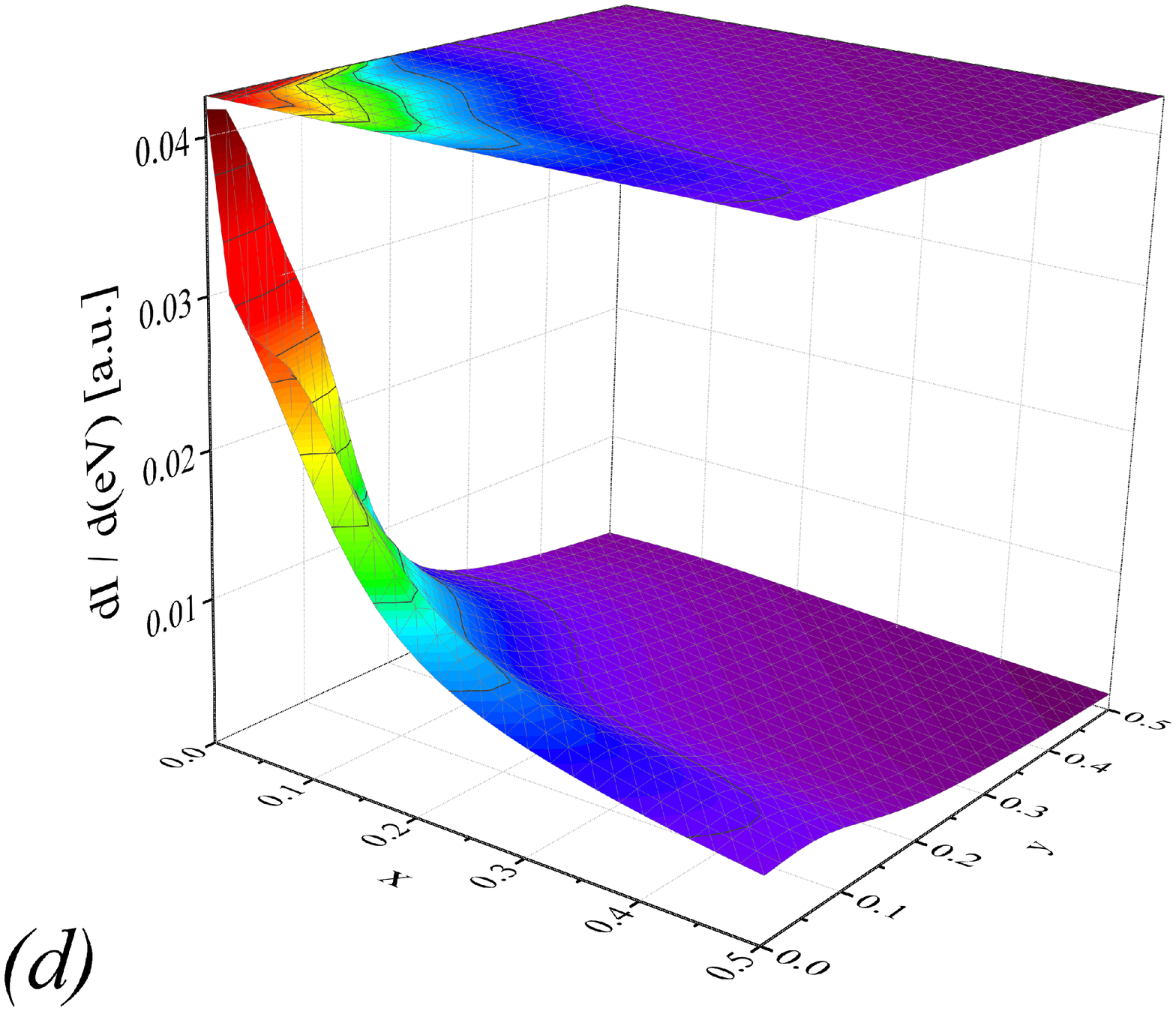}
\includegraphics[width=0.4\textwidth]{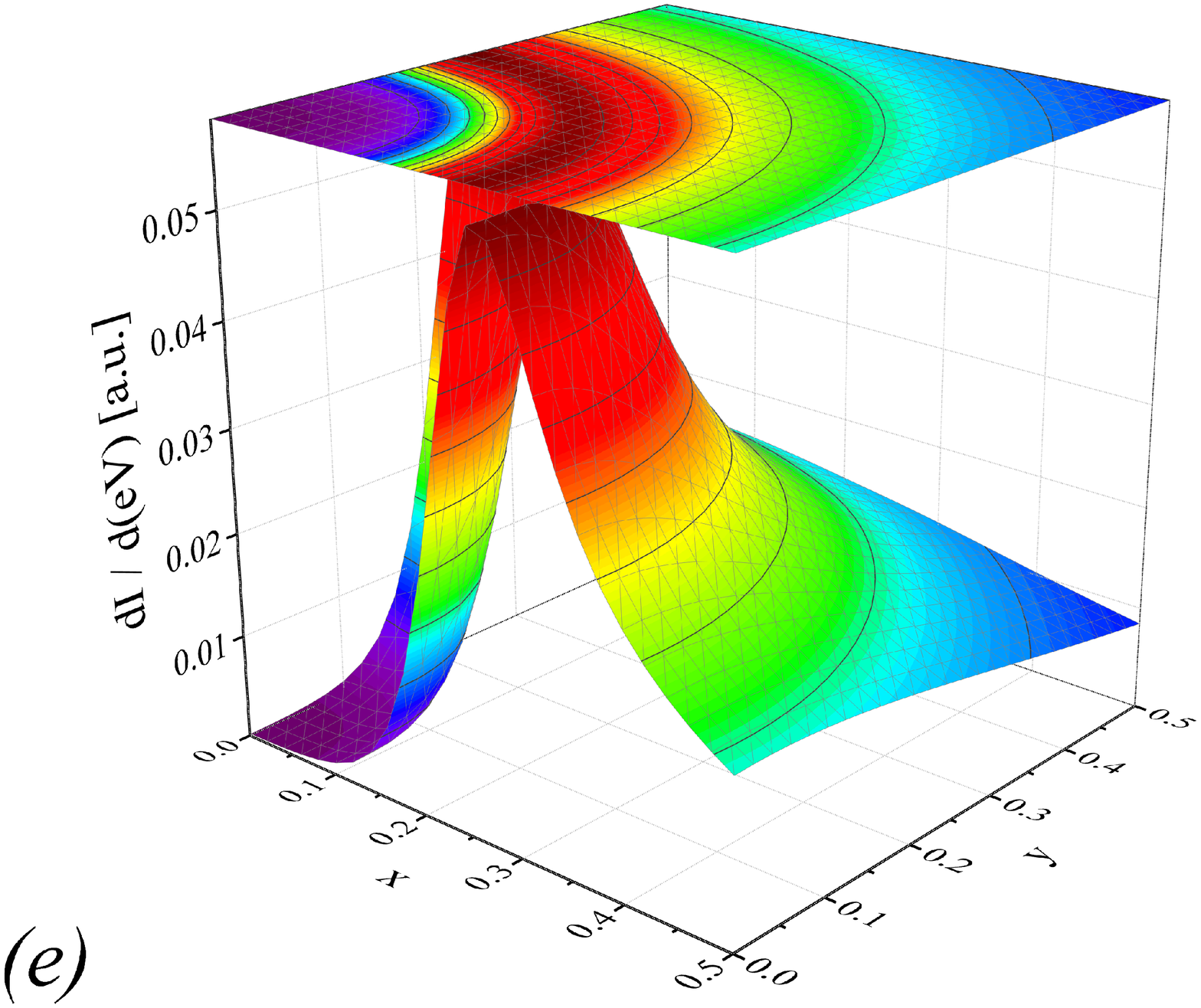}
\includegraphics[width=0.4\textwidth]{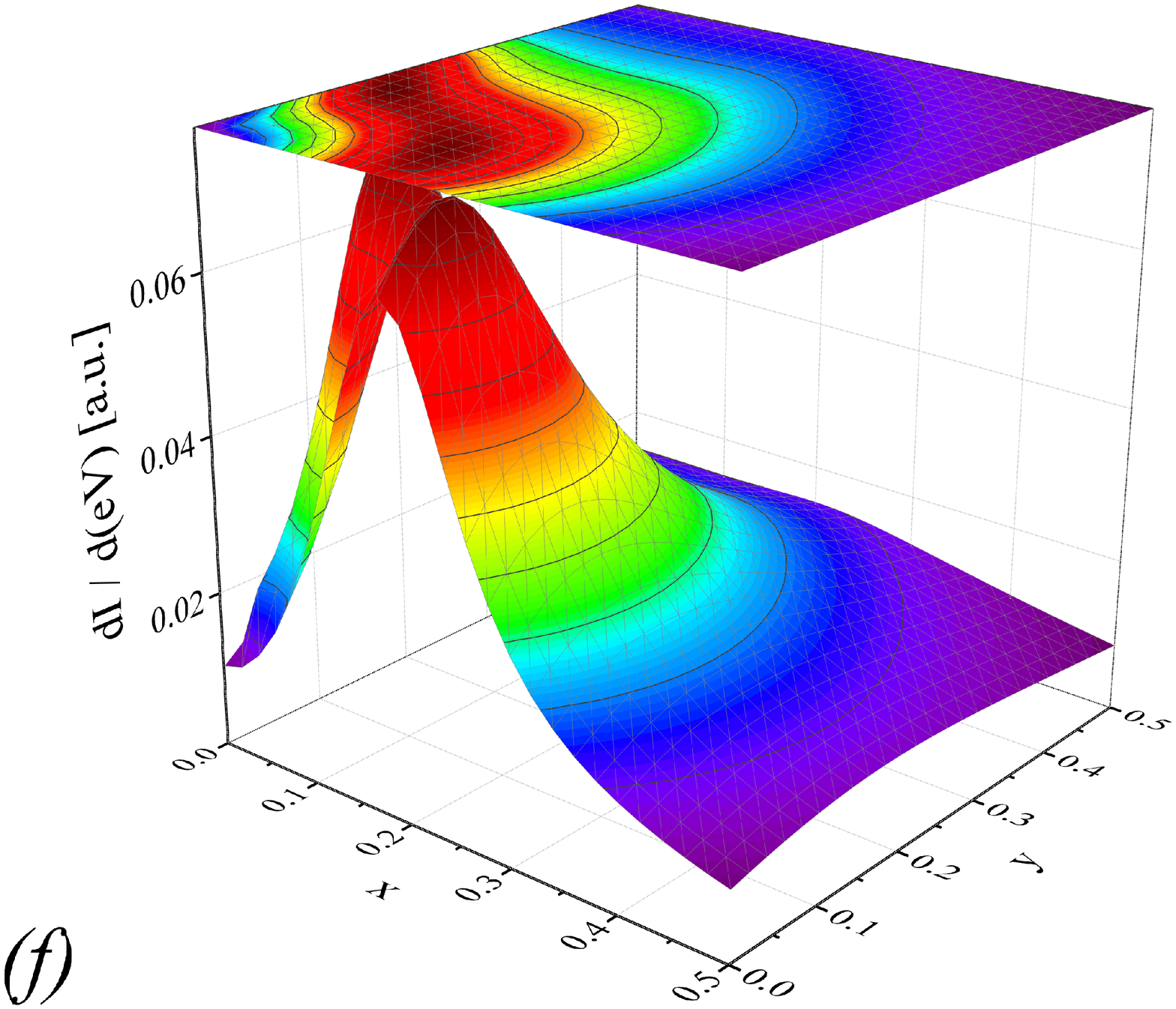}
\caption{(Color online) Evolution of the local differential
conductance $dI/dV( eV,\, x,\, y )$ corresponding to different
bias voltages (($a$),($b$) $eV = 0$; ($c$),($d$) $eV / \Delta_0 =
0.1$; ($e$),($f$) $eV / \Delta_0 = 0.2$)
for different values of the dimensionless barrier
strength $Z$: left column $Z = 0.1$; right column $Z = 0.3$. We
put here $T/\Delta_0=\mathrm{0.02}$.} \label{Fig:8}
\end{figure*}

We have calculated the differential conductance  using
Eqs.~(\ref{eq:LDC1}),(\ref{eq:LDC2}),(\ref{eq:LDC5}) for low
temperature $T/\Delta_0=\mathrm{0.02}$ for different values of the
dimensionless barrier strength $Z$. The typical examples of
dependence of the local differential conductance $d I/ d V$ vs the
bias voltage $eV$ at various distances $r$ from the vortex axis
are shown in Fig.~\ref{Fig:7}. In order to compare our results
with the standard CdGM ones, we present the dependence of the
local $dI/dV$ vs voltage at different distances $r$ from the
Abrikosov vortex axis in the absence of the barrier ($Z=0$). One
can clearly observe the disappearance of the zero bias peak in the
core ($r=0$) and opening of the soft spectral minigap
$\Delta_{Soft}$ caused by the normal scattering at the defect
(Fig.~\ref{Fig:7}($a$)). The barrier results in the anisotropy of
the LDC structure in the plane ($x,\,y$)
(Fig.~\ref{Fig:7}($b,c$)). The anisotropy of the LDC grows when
barrier strength $Z$ increases. Figure~\ref{Fig:8} illustrates the
evolution of the local differential conductance $dI/dV( eV,\, x,\,
y )$ distribution in the plane ($x,\,y$) for several values of the
bias voltage $V$ and dimensionless barrier strength $Z$. In
Fig.~\ref{Fig:8}($a,b$) we can see the spread of the zero bias
peak along the defect which appears to be another hallmark of the
crossover from the Abrikosov to the Josephson vortex type. Due to
the normal reflection of electrons and holes at the defect plane
we get the azimuthal modulation of the LDC developing with the
growth of the barrier strength $Z$.


\section{Summary}\label{sum}

To summarize, we have investigated the transformation of the
subgap spectrum of quasiparticle excitations in the Abrikosov
vortex pinned by the planar defect with a high transparency. We
find that the normal scattering at the defect surface results in
the opening of a soft minigap $\Delta_{Soft}$ in the elementary
excitation spectrum near the Fermi level. The minigap size grows
with the decrease in the transparency of the barrier. The increase
in the resulting soft gap affects the splitting of the zero bias
anomaly in the tunneling spectral characteristics and perturb the
circular symmetry of the LDOS peaks. The normal reflection of
electrons and holes at the defect plane changes the topology of
the isoenergetic orbits in ($\mu-\theta_p$) space. This
topological transition revealing in the specific behavior of the
quantized quasiparticle levels and density of states, can be
considered as a hallmark of the crossover from the Abrikosov to
the Josephson vortex. As a result, there appears a new type of
subgap quasiparticle states gliding along the defect, which reveal
the qualitatively new behavior of discrete spectrum $\varepsilon_n
\sim n^{1/2}$. The hard minigap $\varepsilon_0 \ll \Delta_{Soft}$
in the spectrum of energy levels exceeds noticeably the value of
the CdGM minigap $\hbar \omega_0 \ll \varepsilon_0$. The decrease
in the barrier transparency is accompanied by the increase in the
hard minigap $\varepsilon_0$ in the spectrum which can be observed
in the measurements of the Ohmic and Hall conductivities at finite
frequencies. The basic properties of the vortex such as pinning
and mobility along the defect plane are strongly affected by these
changes in the orbit topology. We have also analyzed the
distinctive features of the quasiparticle density of states, which
accompany the transformation of the subgap quasiparticle spectrum
and the topology of the isoenergetic orbits for an Abrikosov
vortex pinned by a planar defect with a perfect boundary. One can
expect, however, that barrier imperfections and roughness should
result in a partial smearing of both the hard and soft
 gap features similarly to the effect of the point impurity scattering.

Finally, we note that recently the vortices pinned by the defects
are studied as the hosts for the Majorana states in the systems
consisting of a primary superconductor with conventional pairing
and a low dimensional layer with a nontrivial
topology\cite{Rakhmanov-prb11,Ioselevich-prl11,Ioselevich-prb12,Rakhmanov-prb14}.
The isolating inclusions in the vortex core in the primary
superconductor allow to shift the low energy core spectrum from
the Fermi level improving the topological protection of the
Majorana states in the 2D topological superconductor. The vortex
at the planar defect considered in our work can provide a
perspective platform for such states since the hard minigap in the
core can exhibit a strong increase even in the limit of the defect
with high transparency when the shape of the gap inside the vortex
core is only weakly perturbed by the scattering. Another advantage
of this geometry is related to the possibility to move the
vortices along the defects changing, thus, the positions of the
Majorana states in the attached 2D layer without changing the
minigap responsible for the desired topological protection.

\section*{ACKNOWLEDGEMENTS}\label{Acknow}
We thank A.I.Buzdin, Ya.V.Fominov, V.B.Geshkenbein, A.Bezryadin
and A.A.Bespalov for stimulating discussions. This work was
supported in part by the Russian Foundation for Basic Research
under Grants No. 18-42-520037, No. 19-31-51019, the Russian State
Contract No. 0035-2019-0021, and the Foundation for the
Advancement of Theoretical Physics and Mathematics “BASIS” Grant
No. 18-1-2-64-2. The work on the LDOS calculations was supported
by Russian Science Foundation (Grant No. 20-12-00053).


\begin{thebibliography}{0}

\bibitem{Abrikosov57}
A. A. Abrikosov, Sov. Phys. JETP \textbf{5}, 1174 (1957)
[Zh. Eksp. Teor. Fiz. \textbf{32}, 1442 (1957)]. %
%
\bibitem{Barone-Josephson}
A. Barone and G. Paterno, Physics and Applications of the
Josephson Effect (Wiley, New York, 1982).
%
%
\bibitem{Gurevich-PRB92}
A. Gurevich, Phys.\ Rev.\ B \textbf{46}, R3187 (1992).
%
\bibitem{Horide-PRB07}
T. Horide, K. Matsumoto, A. Ichinose, M. Mukaida, Y. Yoshida, S.
Horii, Phys.\ Rev.\ B \textbf{75}, 020504(R) (2007).
%
\bibitem{Horide-PRB08}
T. Horide, K. Matsumoto, Y. Yoshida, M. Mukaida, A. Ichinose, S.
Horii, Phys.\ Rev.\ B \textbf{77}, 132502 (2008).
%
\bibitem{Khlyustikov-Buzdin-AdvPh87}
I. N. Khlyustikov, A. I. Buzdin, Adv. Phys. \textbf{36}, 271
(1987).
%
\bibitem{Gurevich-PRL02}
A. Gurevich, M. S. Rzchowski, G. Daniels, S. Patnaik, B. M.
Hinaus, F. Carillo, F. Tafuri, and D. C. Larbalestier, Phys.\
Rev.\ Lett. \textbf{88}, 097001 (2002).
%
\bibitem{Hilgenkamp-RMP02}
H. Hilgenkamp and J. Mannhart, Rev. Mod. Phys.  \textbf{74}, 485
(2002).
%
\bibitem{Jooss-PRB00}
Ch. Jooss, R. Warthmann and H. Kronmuller, Phys. Rev. B
\textbf{61}, 12433 (2000).
%
\bibitem{Djupmyr-PRB05}
M. Djupmyr, G. Cristiani, H. U. Habermeier, and J. Albrecht, Phys.
Rev. B \textbf{72}, 220507(R) (2005).
%
\bibitem{Tafuri1-RPP05}
F. Tafuri1 and J. R. Kirtley, Rep. Prog. Phys. \textbf{68}, 2573
(2005).
%
\bibitem{Golubov-RMP04}
A. A. Golubov, M. Y. Kupriyanov, , E. Il’ichev, Rev. Mod. Phys.
\textbf{76} 411 (2004).
%
\bibitem{Hess-PRL89}
H.\ F.\ Hess, R.\ B.\ Robinson, R.\ C.\ Dynes, J.\ M.\ Valles,
Jr., and J.\ V.\ Waszczak, Phys.\ Rev.\ Lett.\ \textbf{62}, 214
(1989).
%
\bibitem{Hoogenboom-PRB00}
B.\ W.\ Hoogenboom, M.\ Kugler, B.\ Revaz, I.\ Maggio-Aprile, O.\
Fischer, and Ch.\ Renner, Phys.\ Rev.\ B \textbf{62}, 9179 (2000)
%
\bibitem{Guillamon-PRL08}
I.\ Guillamon, H.\ Suderow, S.\ Vieira, L.\ Cario, P.\ Diener, and
P.\ Rodiere, Phys.\ Rev.\ Lett.\ \textbf{101}, 166407 (2008)
%
\bibitem{Karapetrov-PRL05}
G.\ Karapetrov, J.\ Fedor, M.\ Iavarone, D.\ Rosenmann,
and W.\ K.\ Kwok, %
Phys.\ Rev.\ Lett.\ \textbf{95}, 167002 (2005).
%
\bibitem{Roditchev-NatPh15}
D. Roditchev, C. Brun, L. Serrier-Garcia, et al., Nature Physics
\textbf{11}, 332 (2015).
%
\bibitem{Kopnin-TNSC}
N.\ B.\ Kopnin, {\it Theory of Nonequilibrium Superconductivity}
(Clarendon Press, Oxford, 2001).
%
\bibitem{Guinea-Pogorelov-PRL95}
F. Guinea, Yu. Pogorelov, Phys. Rev. Lett.  \textbf{74}, 462
(1995).
%
\bibitem{Feigelman-Skvortsov-PRL97}
M.V. Feigel’man, M.A. Skvortsov, Phys. Rev. Lett.  \textbf{78},
2640 (1997).
%
\bibitem{Larkin-Ovchinnikov-PRB98}
A.I. Larkin, Yu.N. Ovchinnikov, Phys. Rev. B \textbf{57}, 5457
(1998).
%
\bibitem{Skvortsov-PRB03}
M.A. Skvortsov, D.A. Ivanov, G. Blatter, Phys. Rev. B \textbf{67},
014521 (2003).
%
\bibitem{Melnikov-AVS-JETPL11}
A. S. Mel’nikov and A. V. Samokhvalov, JETP Lett. \textbf{94}, 759
(2011) [Pis'ma v ZhETF \textbf{94}, 823 (2011)].
%
\bibitem{Caroli-PL64}
C.\ Caroli, P.\ G.\ de Gennes, J.\ Matricon, Phys.\ Lett.\
\textbf{9}, 307 (1964).
%
\bibitem{Volovik-Universe} G.\ E.\ Volovik, {\it The Universe in a Helium Droplet},
Clarendon Press, Oxford, 2003.
%
\bibitem{Larkin-PRB98}
A.\ I.\ Larkin and Yu.\ N.\ Ovchinnikov, Phys.\ Rev.\ B
\textbf{57}, 5457 (1998).
%
\bibitem{Tanaka-JJAP95} Y.\ Tanaka, S.\ Kashiwaya, and H.\ Takayanagi, Jpn.\ J.\ Appl.\ Phys.\, Part 1 {\bf 34}, 4566 (1995).
%
\bibitem{Eschrig} M.\ Eschrig, D.\ Rainer, and J.\ A.\ Sauls: in {\it Vortices in unconventional superconductors and
superfluids}, ed. R.P. Huebener, N. Schopohl and G.E. Volovik
(Springer Verlag, Berlin, 2001), preprint cond-mat/0106546.
%
\bibitem{Melnikov-Samokhvalov-PRB09}
A.S. Mel'nikov, A.V. Samokhvalov, M.N. Zubarev, Phys. Rev. B
\textbf{79}, 134529 (2009).
%
\bibitem{Rosenstein-PRB11_microwave-absorp}
B. Rosenstein, I. Shapiro, E. Deutch, B.Ya. Shapiro, Phys. Rev. B
\textbf{84}, 134521 (2011).
%
\bibitem{Melnikov-Samokhvalov-JETPL15_review}
A.S. Mel'nikov, A.V.Samokhvalov, V.L. Vadimov, Письма в ЖЭТФ
\textbf{102}, 886 (2015).
%
\bibitem{Vadimov-Melnikov-JLTP16_chiral}
V.L. Vadimov, A.S. Mel'nikov, J. Low Temp. Phys. \textbf{183}, 342
(2016).
%
\bibitem{LL-IX-2}
L.D. Landau, L.P. Pitaevskii, {\it Statistical Physics, Part 2},
(Oxford: Pergamon, Ch.5, 1980).
%
\bibitem{Kopnin-PRL05}
N. B. Kopnin, A. S. Mel’nikov, V. I. Pozdnyakova, D. A. Ryzhov, I.
A. Shereshevskii, and V. M. Vinokur, Phys. Rev. Lett. \textbf{95},
197002 (2005).
%
\bibitem{Kopnin-PRB07}
N.\ B.\ Kopnin, A.\ S.\ Melnikov, V.\ I.\ Pozdnyakova, D.\ A.\
Ryzhov, I.\ A.\ Shereshevskii, and V.\ M.\ Vinokur, Phys.\ Rev.\ B
\textbf{75}, 024514 (2007).
%
\bibitem{Melnikov-PRB08}
A.\ S.\ Mel'nikov, D.\ A.\ Ryzhov, and M.\ A.\ Silaev, Phys.\
Rev.\ B \textbf{78}, 064513 (2008).
%
\bibitem{Melnikov-PRB09-LDOS-mesa}
A.\ S.\ Mel'nikov, D.\ A.\ Ryzhov, and M.\ A.\ Silaev, Phys.\
Rev.\ B \textbf{79}, 134521 (2009).
%
\bibitem{lifshits}
I. M. Lifshits, Zh.\ Eksp.\ Teor.\ Fiz.\ {\bf 38}, 1569 (1960)
[Sov.\ Phys.\ JETP {\bf 11}, 1130 (1960)].
%
\bibitem{blanter}
Y.\ M.\ Blanter, M.\ I.\ Kaganov, A.\ V.\ Pantsulaya,
 A.\ A.\ Varlamov, Phys.\ Reports {\bf 245}, 159 (1994).
%
\bibitem{Volovik-gapless}
G.\ E.\ Volovik, Pis'ma Zh.\ Eksp.\ Teor.\ Fiz.\ \textbf{49}, 343
(1989) [JETP Lett.\ \textbf{},  (1989)].
%
 \bibitem{mel-silaev}
 S. Mel'nikov and M. A. Silaev, Pis'ma Zh. Eksp. Teor. Fiz. 83,
675 (2006) [JETP Lett. 83, 578 (2006)].
%
\bibitem{Beenakker}
C.\ W.\ J.\ Beenakker and H.\ van Houten, Phys.\ Rev.\ Lett.\ {\bf
66}, 3056, (1991); C.\ W.\ J.\ Beenakker, Phys.\ Rev.\ Lett.\ {\bf
67}, 3836, (1991).
%
\bibitem{Kopnin-JETPL96-BohrSomm}
N. B. Kopnin and G. E. Volovik, Pis’ma Zh. Eksp. Teor. Fiz.
\textbf{64}, 641 (1996) [JETP Lett. \textbf{64}, 690 (1996)]; N.
B. Kopnin, Phys. Rev. B \textbf{57}, 11775 (1998).
%
\bibitem{Kopnin-PRL97-BohrSomm}
N. B. Kopnin and G. E. Volovik, Phys. Rev. Lett. \textbf{79}, 1377
(1997).
%
\bibitem{graphene1}
D. N. Basov, M. M. Fogler, A. Lanzara, Feng Wang, Yuanbo Zhang,
Rev. Mod. Phys. \textbf{86}, 959 (2014).
\bibitem{graphene2}
A. H. Castro Neto, F. Guinea, N. M. R. Peres, K. S. Novoselov and
A. K. Geim,Rev. Mod. Phys. \textbf{81}, 109 (2009).
%
\bibitem{Janko-prl99}
B. Janko, Phys. Rev. Lett. \textbf{82}, 4703 (1999).
%
\bibitem{Blatter-RMP94}
G.\ Blatter, M.\ V.\ Feigel'man, V.\ B.\ Geshkenbein, A.\ I.\ Larkin, V.\ M.\ Vinokur, %
Rev. Mod. Phys. \textbf{66}, 1125 (1994).
%
\bibitem{BTK-PRB82}
G. E. Blonder, M. Tinkham, and T. M. Klapwijk, Phys. Rev. B
\textbf{25}, 4515 (1982).
%
\bibitem{Kopnin-PRB03}
N. B. Kopnin, A. S. Mel’nikov, and V. M. Vinokur, Phys. Rev. B
\textbf{68}, 054528 (2003).
%
\bibitem{Rakhmanov-prb11}
A. L. Rakhmanov, A. V. Rozhkov, and Franco Nori, Phys. Rev. B
\textbf{84}, 075141 (2011).
%
\bibitem{Ioselevich-prl11}
 P. A. Ioselevich and M. V. Feigel’man, Phys. Rev. Lett.
\textbf{106}, 077003 (2011).
%
\bibitem{Ioselevich-prb12}
P. A. Ioselevich, P. M. Ostrovsky, and M. V. Feigel’man, Phys.
Rev. B \textbf{86}, 035441 (2012).
%
\bibitem{Rakhmanov-prb14}
R. S. Akzyanov, A. V. Rozhkov, A. L. Rakhmanov, and Franco Nori,
Phys. Rev. B \textbf{89}, 085409 (2014).

\end{thebibliography}
\end{document}